\documentclass[aps, pre, reprint, superscriptaddress, longbibliography]{revtex4-1}
\usepackage[dvipdfmx]{graphicx}
\usepackage{amsmath}
\usepackage{amssymb}
\usepackage{dcolumn}
\usepackage{color}
\usepackage{bm}
\usepackage{braket}
\usepackage{txfonts}

\begin{document}
\title{Machine learning analysis of dimensional reduction conjecture for nonequilibrium Berezinskii--Kosterlitz--Thouless transition in three dimensions}
\author{Taiki Haga}
\email[]{taiki.haga@omu.ac.jp}
\affiliation{Department of Physics and Electronics, Osaka Metropolitan University, Sakai-shi, Osaka 599-8531, Japan}
\date{\today}

\begin{abstract}
We investigate the recently proposed dimensional reduction conjecture in driven disordered systems using a machine learning technique.
The conjecture states that a static snapshot of a disordered system driven at a constant velocity is equal to a space-time trajectory of its lower-dimensional pure counterpart.
This suggests that the three-dimensional random field XY model exhibits the Berezinskii--Kosterlitz--Thouless transition when driven out of equilibrium.
To verify the conjecture directly by observing configurations of the system, we utilize the capacity of neural networks to detect subtle features of images.
Specifically, we train neural networks to differentiate snapshots of the three-dimensional driven random field XY model from space-time trajectories of the two-dimensional pure XY model.
Our results demonstrate that the network cannot distinguish between the two, confirming the dimensional reduction conjecture.
\end{abstract}

\maketitle

\section{Introduction}

The large-scale behavior of classical and quantum many-body systems in nonequilibrium steady states has been a central research topic in modern statistical physics.
The absence of the detailed balance due to nonequilibrium driving results in unique behaviors that are not found in thermal equilibrium, such as long-range order in two-dimensional (2D) systems with continuous symmetry \cite{Toner-95, Nakano-21}, absorbing-state phase transitions \cite{Hinrichsen-00, Takeuchi-07, Sano-16}, and motility-induced phase separation \cite{Cates-15, Tailleur-08, Fily-12, Redner-13}.
Recent advancements in experiments, from ultracold atomic gases to biological systems, have heightened interest in this expansive subject.

\begin{figure}[b]
\centering
\includegraphics[width=8.6cm]{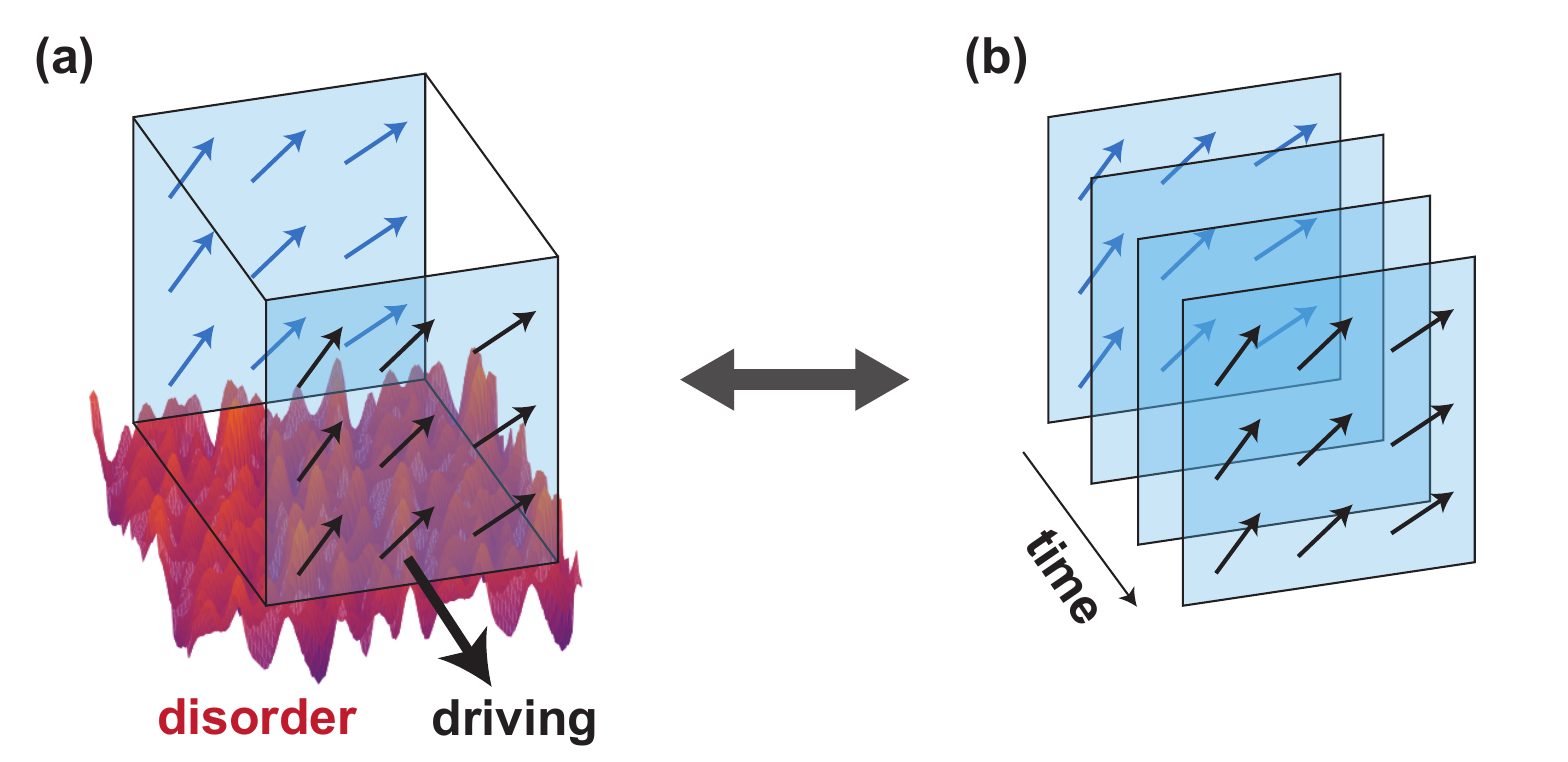}
\caption{Schematic illustration of the dimensional reduction conjecture. It predicts that a snapshot of $D$-dimensional disordered systems driven at a constant velocity [panel (a)] is equivalent to a space-time trajectory of the $(D-1)$-dimensional pure system [panel (b)].}
\label{fig_DR_concept}
\end{figure}

Recent studies suggest that when the three-dimensional (3D) XY model with a random field is driven at a constant velocity, it shows the Berezinskii--Kosterlitz--Thouless (BKT) transition \cite{Haga-15, Haga-18}.
This finding is striking because, under thermal equilibrium, the 3D random field XY model remains disordered and does not exhibit any phase transition \cite{Imry-75, Aizenman-89}.
The principle behind the 3D nonequilibrium BKT transition is the dimensional reduction conjecture \cite{Haga-19, Haga-17}.
In simple terms, it proposes that {\it a static snapshot of a $D$-dimensional disordered system driven at a constant velocity is equal to a space-time trajectory of its $(D-1)$-dimensional pure counterpart} (see Fig.~\ref{fig_DR_concept} for a schematic illustration).
Using this idea, the 3D driven random field XY model equates to the 2D pure XY model, leading to the expectation of the BKT transition in the 3D model. 
However, it remains unclear when and under which circumstances this dimensional reduction holds true, as there are simple counterexamples (see Appendix \ref{sec:failure_of_dimensional_reduction}).

This study aims to test the dimensional reduction conjecture using machine learning.
Neural networks excel at identifying subtle data patterns that humans cannot capture.
In statistical physics, they have been employed to identify and categorize thermodynamic phases in classical spin systems \cite{Carrasquilla-17, Singh-17, Wetzel-17, Wang-17, Beach-18, Greitemann-19}, glassy systems \cite{Schoenholz-17, Bapst-20, Boattini-20, Jin-21, Oyama-23}, and quantum many-body systems \cite{Nieuwenburg-17, Broecker-17, Venderley-18, Hsu-18, Huembeli-19, Dong-19, Kottmann-20}.
Our goal with the neural network is to compare the structures of a driven disordered system with its simpler, lower-dimensional pure version. 
To do this, we train convolutional and fully-connected neural networks to distinguish between snapshots of the 3D driven random field XY model (DRFXYM) and space-time trajectories of the 2D pure XY model.
This is a standard binary classification task (see Fig.~\ref{fig_neural_network}).
We demonstrate that the classification accuracy of the trained network is no better than a random guess, suggesting that the two systems are indistinguishable. 

This paper is organized as follows:
In Sec.~\ref{sec:model}, we introduce a prototypical model for driven phase ordering systems with disorder, referred to as the driven random field $O(N)$ model.
From a naive argument, we propose the dimensional reduction conjecture for this model.
The main discussion will focus on the 3D DRFXYM ($N=2$).
In Sec.~\ref{sec:helicity_modulus}, before exploring the machine learning verification of dimensional reduction, we highlight characteristics of the 3D BKT transition.
Specifically, we compute the helicity modulus for the 3D DRFXYM, revealing a jump at a certain critical disorder strength. 
Additionally, the jump amplitude of the helicity modulus satisfies the universal jump relation given by Eq.~\eqref{hm_jump_3D}.
Sec.~\ref{sec:test_of_dimensional_reduction} is devoted to our machine learning examination of the dimensional reduction conjecture.
Our findings indicate that the task of distinguishing between the 3D DRFXYM and the 2D pure XY model becomes challenging when the parameters of both models match the anticipated values for model mapping of the dimensional reduction.
We summarize our conclusions in Sec.~\ref{sec:conclusion}.
In Appendix \ref{sec:spin_wave_approximation}, we review the analysis of the spin-wave approximation for the 2D XY model and the 3D DRFXYM.
Appendix \ref{sec:simulation} provides details of our numerical simulation procedures.
In Appendix \ref{sec:coarse_graining_procedure}, we present details of coarse-graining procedures applied to data.
Appendix \ref{sec:reproducibility} discusses the reproducibility of the neural network results.
In Appendix \ref{sec:failure_of_dimensional_reduction}, we discuss a counterexample of the dimensional reduction.
Specifically, we demonstrate that the 2D driven random field Ising model is {\it not} identical to the 1D pure Ising model, emphasizing the intricate nature of the dimensional reduction conjecture.

\section{Model and dimensional reduction}
\label{sec:model}

We introduce a model of driven phase ordering systems with disorder and address the dimensional reduction  conjecture.
Let us consider an $N$-component real vector field given by $\boldsymbol{\phi}(\mathbf{r})=(\phi_1(\boldsymbol{r}), ..., \phi_N(\boldsymbol{r}))$.
The Hamiltonian for the $D$-dimensional $O(N)$ model with a quenched random field $\boldsymbol{h}(\boldsymbol{r})=(h_1(\boldsymbol{r}), ..., h_N(\boldsymbol{r}))$ is
\begin{equation}
H[\boldsymbol{\phi}; \boldsymbol{h}] = \int d^D \boldsymbol{r} \left[ \frac{1}{2} \sum_{i=1}^N |\nabla \phi_i|^2 + U(\rho) - \boldsymbol{h} \cdot \boldsymbol{\phi} \right],
\label{Hamiltonian}
\end{equation}
where $U(\rho) = (g/2) (\rho - 1/2)^2$ denotes the local potential with field amplitude $\rho = |\boldsymbol{\phi}|^2/2$.
The random field $\boldsymbol{h}(\boldsymbol{r})$ obeys a Gaussian distribution with zero mean satisfying
\begin{equation}
\langle h_i(\boldsymbol{r}) h_j(\boldsymbol{r}') \rangle = \Delta^2 \delta_{ij} \delta(\boldsymbol{r}-\boldsymbol{r}'),
\label{random_field_correlation}
\end{equation}
where $\langle ... \rangle$ denotes averaging over disorder realizations and $\Delta$ is a parameter characterizing the intensity of the random field.
Equation \eqref{random_field_correlation} implies that the correlation of the random field is short-range.

The dynamics of the model is described by
\begin{equation}
\partial_t \phi_i + \Lambda[\phi_i] = - \frac{\delta H[\boldsymbol{\phi}; \boldsymbol{h}] }{\delta \phi_i} + \xi_i,
\label{model_equation_Lambda}
\end{equation}
where $\xi_i(\boldsymbol{r}, t)$ represents the thermal noise satisfying
\begin{equation}
\langle \xi_i(\boldsymbol{r}, t) \xi_j(\boldsymbol{r}', t') \rangle_T = 2 T \delta_{ij} \delta(\boldsymbol{r}-\boldsymbol{r}') \delta(t - t').
\label{thermal_noise_xi}
\end{equation}
In this, $\langle ... \rangle_T$ denotes averaging over the thermal noise.
The term $\Lambda[\phi_i]$ in Eq.~\eqref{model_equation_Lambda} signifies the effect of nonequilibrium driving.
For $\Lambda[\phi_i]=0$ and $\boldsymbol{h}=0$, Eq.~\eqref{model_equation_Lambda} corresponds to the conventional dynamics of phase ordering systems with a non-conservative order parameter, known as ``model A" \cite{Hohenberg-77}.
We postulate that $\Lambda[\phi]$ satisfies the following conditions:
\begin{itemize}
\item Non-potentiality: There is no potential functional $V[\phi]$ such that $\Lambda[\phi] = \delta V[\phi] / \delta \phi$.
\item Locality: $\Lambda[\phi]$ is a function of $\phi$ and its spatial derivatives at the same space-time point.
\item Symmetry: $\Lambda[\phi]$ is transformed in the same way as $\phi$ by the transformation corresponding to the $Z_2$ or $O(N)$ symmetry.
\item Linearity: $\Lambda[\phi]$ is linear with respect to $\phi$.
\end{itemize}
The simplest choice is $\Lambda[\phi_i]=(\boldsymbol{v} \cdot \nabla) \phi_i$ with a constant vector $\boldsymbol{v}$.
Therefore, the dynamics simplifies to:
\begin{equation}
\partial_t \phi_i + v \partial_x \phi_i = - \frac{\delta H[\boldsymbol{\phi}; \boldsymbol{h}] }{\delta \phi_i} + \xi_i.
\label{model_equation}
\end{equation}
Hereafter, we refer to this as the driven random field $O(N)$ model [DRF$O(N)$M], which has been firstly introduced in Ref.~\cite{Haga-15}.

The DRF$O(N)$M characterizes the relaxation dynamics of ordered systems flowing in a random environment, exemplified by liquid crystals flowing through porous media. 
Recent interest has surged in the dynamics of liquid crystals within complex geometries, driven by both fundamental research interest and industrial applications \cite{Araki-12}.
In porous media, the irregular surface of the substrate induces random anchoring that breaks the symmetry, analogous to the random field in the $O(N)$ model.

We discuss the dimensional reduction conjecture for the DRF$O(N)$M \cite{Haga-17, Haga-19}.
At zero temperature $T=0$ (without the thermal noise $\xi_i$), Eq.~\eqref{model_equation} is written as
\begin{equation}
\partial_t \phi_i + v \partial_x \phi_i = \left[ \nabla^2 - U'(\rho) \right] \phi_i + h_i.
\end{equation}
We assume that, after a sufficient long time, the system relaxes to a steady state, which satisfies
\begin{equation}
v \partial_x \phi_i = \left[ \nabla^2 - U'(\rho) \right] \phi_i + h_i.
\label{steady_state_equation_1}
\end{equation}
In the large length scale, the longitudinal elastic term $\partial_x^2 \phi_i$ is negligible compared to the advection term $v \partial_x \phi_i$. 
Thus, Eq.~\eqref{steady_state_equation_1} can be rewritten as
\begin{equation}
v \partial_x \phi_i(x, \boldsymbol{r}_\perp) = \left[ \nabla_\perp^2 - U'(\rho(x, \boldsymbol{r}_\perp)) \right] \phi_i(x, \boldsymbol{r}_\perp) + h_i(x, \boldsymbol{r}_\perp),
\label{steady_state_equation_2}
\end{equation}
where $\nabla_\perp$ represents the spatial derivative with respect to the transverse directions and $\boldsymbol{r}_\perp$ represents the transverse coordinate.
If the coordinate $x$ is considered to be a fictitious time and $h_i(x, \boldsymbol{r}_\perp)$ as thermal noise, Eq.~\eqref{steady_state_equation_2} is equal to the dynamical equation for the $(D-1)$-dimensional pure $O(N)$ model with an effective temperature 
\begin{equation}
T_\mathrm{eff} = \frac{\Delta^2}{2v}.
\label{effective_temperature}
\end{equation}
This naive argument concludes that a static snapshot of the $D$-dimensional DRF$O(N)$M at zero temperature is identical to a space-time trajectory of the $(D-1)$-dimensional pure $O(N)$ model at finite temperature. 
We refer to this as the dimensional reduction conjecture, which is illustrated in Fig.~\ref{fig_DR_concept}.

We now turn our attention to the case where $N=2$, the driven random field XY model (DRFXYM).
If the dimensional reduction conjecture is correct, the 3D DRFXYM corresponds to the usual 2D XY model.
Given that the 2D XY model undergoes the BKT transition at a critical temperature $T_c$, it is anticipated that the 3D DRFXYM will similarly exhibit a transition at a critical disorder $\Delta_c$.
From Eq.~\eqref{effective_temperature}, the critical disorder can be expressed as
\begin{equation}
\Delta_c=\sqrt{2v T_c}.
\label{Delta_c_DR}
\end{equation}

We note that the heuristic argument leading to the dimensional reduction rests on nontrivial assumptions.
Firstly, it is not trivial whether the system always relaxes to a steady state after a sufficiently long time.
The system might exhibit chaotic fluctuations indefinitely, even if there is no thermal noise.
Secondly, typical steady states satisfying Eq.~\eqref{steady_state_equation_2} are not guaranteed to be the typical dynamical solutions of the lower dimensional counterpart.
More precisely, the realization probability of steady states satisfying Eq.~\eqref{steady_state_equation_2} might be quite different from that of space-time trajectories of lower dimensional counterpart at thermal equilibrium.
The aforementioned argument does not provide clarity on the comparative realizability of one steady state over another.
Given these intricacies, the applicability of the dimensional reduction conjecture to a driven disordered system remains a nontrivial issue.
In the main body of this paper, we demonstrate that dimensional reduction is valid for DRFXYM.
However, as discussed in Appendix \ref{sec:failure_of_dimensional_reduction}, the dimensional reduction does not hold for the case of $N=1$, i.e., the driven random field Ising model.

We briefly remark on the dimensional reduction conjecture for disordered systems in thermal equilibrium, a concept established in the 1970s.
This predicts that the large-scale physics of $D$-dimensional disordered systems is the same as that of the $(D-2)$-dimensional counterparts without disorder \cite{Aharony-76, Young-77, Grinstein-76, Parisi-79}.
Well-studied examples of such systems include elastic manifolds in random media \cite{Fisher-86, Balents-93, Giamarchi-95, LeDoussal-03, LeDoussal-04}, the random field Ising model \cite{Imbrie-84, Bricmont-87}, and the random field and random anisotropy $O(N)$ models \cite{Fisher-85, Feldman-02, Tarjus-08, Tissier-08}.
Nonetheless, it is recognized that the dimensional reduction can break down due to nonperturbative effects linked to the existence of numerous energy landscape local minima \cite{Fisher-86, Balents-93, Giamarchi-95, LeDoussal-03, LeDoussal-04, Imbrie-84, Bricmont-87, Fisher-85, Feldman-02, Tarjus-08, Tissier-08}.
Our dimensional reduction conjecture discussed in this paper can be considered as a nonequilibrium counterpart of this well-established concept.
In nonequilibrium cases, the similar nonperturbative effects arising from an abundance of steady states could undermine the validity of the dimensional reduction \cite{Haga-17, Haga-19}.

\section{Universal jump of helicity modulus}
\label{sec:helicity_modulus}

\begin{figure*}
\centering
\includegraphics[width=\textwidth]{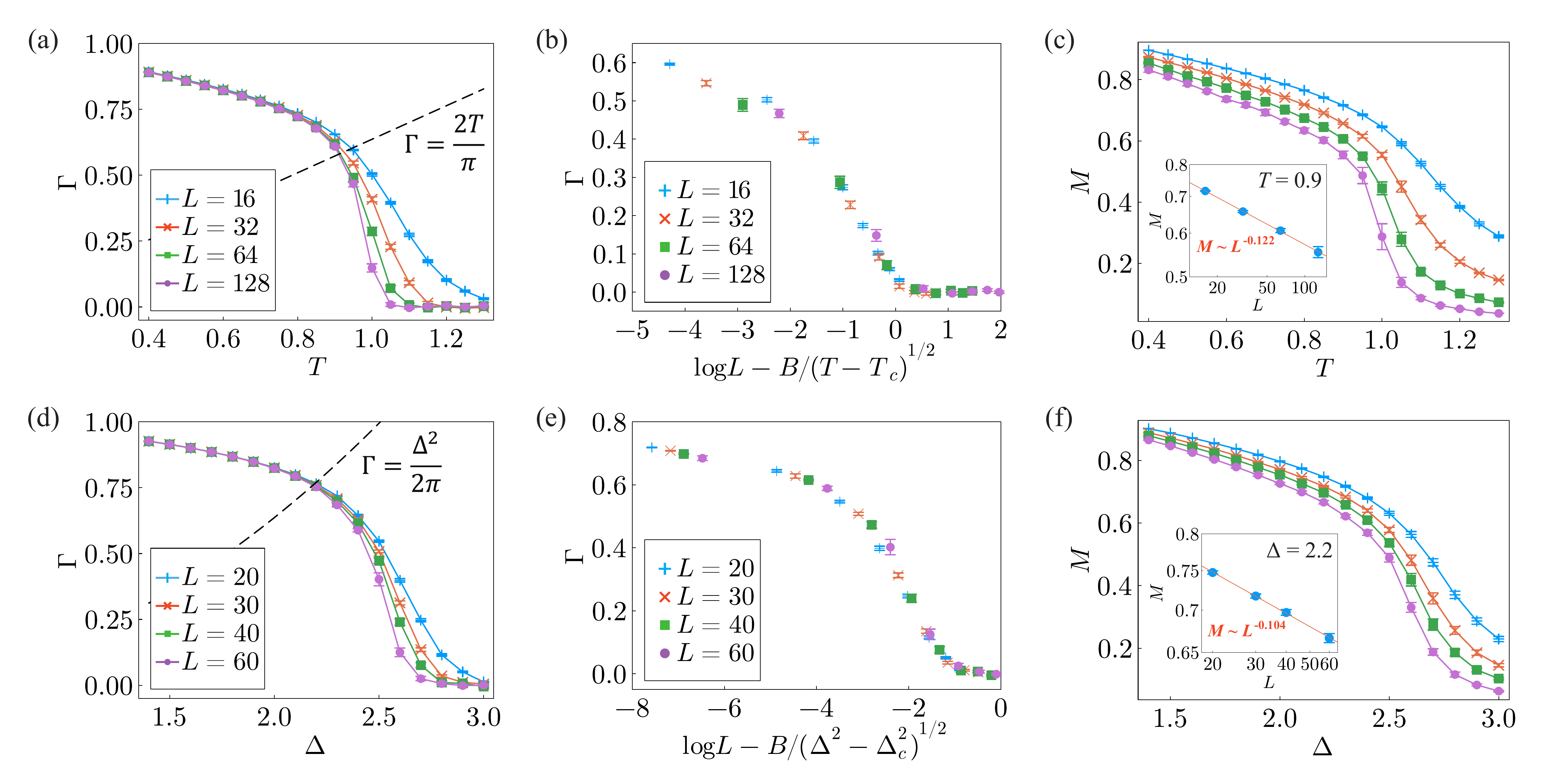}
\caption{Helicity modulus $\Gamma$ and magnetization $M$ for the 2D XY model [panels (a)-(c)] and the 3D DRFXYM [panels (d)-(f)].
(a) Helicity modulus $\Gamma$ for the 2D XY model as a function of temperature with systems sizes $L=16, 32, 64, 128$.
The dashed line represents $\Gamma = 2T/\pi$.
(b) Finite-size scaling plot of $\Gamma$ as a function of $\log L - B/(T-T_c)^{1/2}$.
The scaling parameters are $T_c = 0.89$ and $B = 1.73$.
(c) Magnetization $M$ for the 2D XY model with systems sizes $L=16, 32, 64, 128$ from top to bottom.
The inset shows the system size dependence of $M$ at the transition point, shown in log-scale on the both axes.
(d) Helicity modulus $\Gamma$ for the 3D DRFXYM as a function of disorder $\Delta$ with systems sizes $L=20, 30, 40, 60$.
The driving velocity is $v=2$.
The dashed curve represents $\Gamma = \Delta^2/(v \pi)$.
(e) Finite-size scaling plot of $\Gamma$ as a function of $\log L - B/(\Delta^2-\Delta_c^2)^{1/2}$.
The scaling parameters are $\Delta_c = 2.17$ and $B = 8.06$.
(f) Magnetization $M$ for the 3D DRFXYM with systems sizes $L=20, 30, 40, 60$ from top to bottom.
The inset shows the system size dependence of $M$ at the transition point, shown in log-scale on the both axes.}
\label{fig_helicity_modulus}
\end{figure*}

\begin{figure}
\centering
\includegraphics[width=8.6cm]{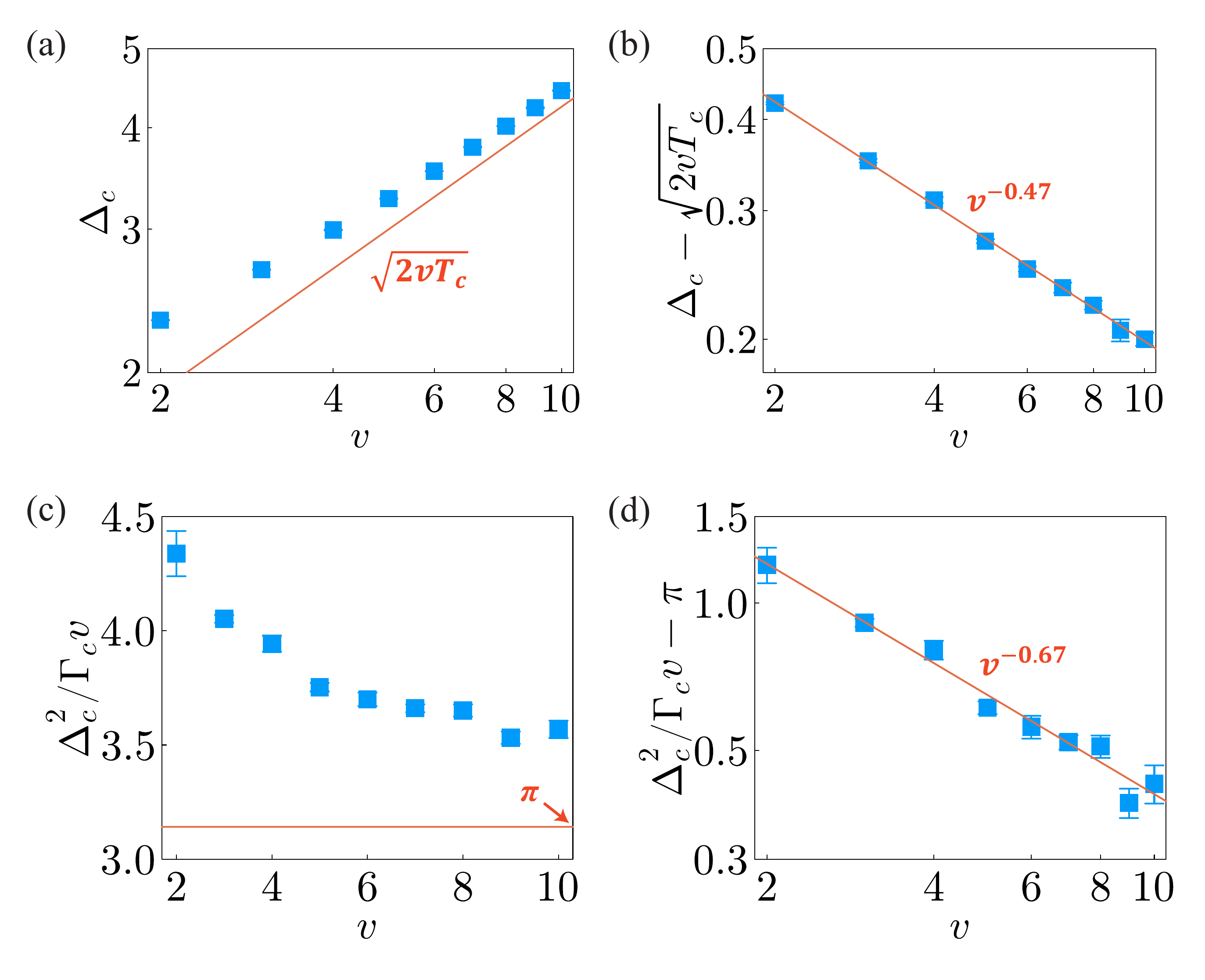}
\caption{Verification of the universal jump relation for the helicity modulus \eqref{hm_jump_3D}.
(a) The critical disorder intensity $\Delta_c$ as a function of the driving velocity $v$, shown in log-scale on the both axes. 
The solid line represents $\sqrt{2vT_c}$, where $T_c$ is the critical temperature of the 2D XY model [see Eq.~\eqref{Delta_c_DR}]. 
(b) $\Delta_c-\sqrt{2vT_c}$ as a function of $v$, shown in log-scale on the both axes. 
The solid line represents the least-square fitting by $av^{-b}$, where the exponent is estimated as $b=0.47$. 
(c) $\Delta_c^2/\Gamma_c v$ as a function of the driving velocity $v$, where $\Gamma_c$ is the jump of the helicity modulus at the transition $\Delta = \Delta_c$. 
The solid line represents the universal jump relation \eqref{hm_jump_3D}. 
(d) $\Delta_c^2/\Gamma_c v-\pi$ as a function of $v$, shown in log-scale on the both axes.
The solid line represents the least-square fitting by $av^{-b}$, where the exponent is estimated as $b=0.67$.}
\label{fig_hm_jump}
\end{figure}

Before proceeding to the test of the dimensional reduction conjecture using machine learning, we first discuss the basic characteristics of the BKT transition in the 3D DRFXYM.
The conventional BKT transition is signified by a discontinuous jump in the helicity modulus \cite{Nelson-77, Ohta-79}.
In this section, we examine the helicity modulus of the 3D DRFXYM, demonstrating that it similarly exhibits a jump at the transition point.

\subsection{2D XY model}

We begin by examining the standard 2D XY model as our reference system.
The Hamiltonian is given by
\begin{equation}
H = - \sum_{\langle i j \rangle} \cos (\theta_i - \theta_j),
\end{equation}
where $\theta_i$ is the angle of the spin at site $i$ in a square lattice of length $L$, and $\langle i j \rangle$ denotes a nearest-neighboring pair.

A well-understood aspect of this model is the behavior of the helicity modulus $\Gamma$.
The helicity modulus $\Gamma$ is determined by measuring the response force to the twist of the spins.
Suppose twisting the spins at the left boundary of a 2D square simulation box of length $L$ by a slight angle $\theta$ with respect to the right boundary.
The helicity modulus $\Gamma$ is then expressed as:
\begin{equation}
	\Gamma = \lim_{\theta \to 0} \frac{1}{\theta} \left\langle \frac{\partial H}{\partial \theta} \right\rangle.
	\label{Gamma_2D_def}
\end{equation}
At the critical temperature $T_c$, $\Gamma$ displays a sharp drop to zero.
The amplitude of this jump in $\Gamma$ satisfies the universal jump relation \cite{Nelson-77, Ohta-79}:
\begin{equation}
\frac{\Gamma_c}{2T_c} = \frac{1}{\pi}.
\label{hm_jump_2D}
\end{equation}

We conduct Monte Carlo simulations over $10^5 L^2$ steps, and the time average for the helicity modulus $\Gamma$ is computed, after an initial transient regime to ensure system equilibration.
Additionally, these measurements are ensemble-averaged over 50 to 100 different initial states to improve statistical reliability and reduce sampling errors.
In Fig.~\ref{fig_helicity_modulus}(a), we show $\Gamma$ as a function of temperature for various system sizes $L$.
For temperatures below $T_c \simeq 0.9$, the value of $\Gamma$ remains unaffected by changes in $L$. 
However, for temperatures above this critical point, $\Gamma$ diminishes to zero as $L$ increases.
The solid line in Fig.~\ref{fig_helicity_modulus}(a) represents $\Gamma = 2T/\pi$.
Using Eq.~\eqref{hm_jump_2D}, this line intersects with the $\Gamma$ curve right at the critical temperature.

To deal with the finite-size effect, it is important to point out that, for different system sizes $L$, the values of $\Gamma$ should collapse to a single curve when plotted against the ratio $L/\xi(T)$, where $\xi(T)$ represents the correlation length \cite{Schultka-94}.
Close to the BKT transition, the correlation length $\xi(T)$ behaves as
\begin{equation}
\xi(T) \propto \exp[B/(T-T_c)^{1/2}],
\label{xi_2D}
\end{equation}
where $B$ is a non-universal constant.
With this in mind, in Fig.~\ref{fig_helicity_modulus}(b), we plot $\Gamma$ as a function of $\log[L/\xi(T)]=\log L - B/(T-T_c)^{1/2}$.
We can observe data collapse with parameters $T_c = 0.89$ and $B = 1.73$.

The BKT transition can also be characterized by the system size dependence of the magnetization.
The magnetization per spin is given by
\begin{equation}
\hat{\mathcal{M}} = \frac{1}{L^2} \int d^2 \mathbf{r} \boldsymbol{\phi}(\mathbf{r}),
\end{equation}
where its average is noted to vanish, $\langle \hat{\mathcal{M}} \rangle=0$.
Consequently, we consider its standard deviation,
\begin{equation}
M := \langle | \hat{\mathcal{M}} |^2 \rangle^{1/2} = \frac{1}{L^2} \left[ \int d^2 \mathbf{r} d^2 \mathbf{r}' \langle \boldsymbol{\phi}(\mathbf{r}) \cdot \boldsymbol{\phi}(\mathbf{r}') \rangle \right]^{1/2}.
\label{M_2D_def}
\end{equation}
Below the critical temperature $T_c \simeq 0.9$, the spin correlation function exhibits a power law decay:
\begin{equation}
\langle \boldsymbol{\phi}(\boldsymbol{r}) \cdot \boldsymbol{\phi}(\boldsymbol{r}') \rangle \propto |\boldsymbol{r}-\boldsymbol{r}'|^{-\eta_\mathrm{2D}(T)},
\label{spin_correlation_2D}
\end{equation}
where $\eta_\mathrm{2D}(T)$ is a temperature-dependent exponent.
With the spin-wave approximation, which is reviewed in Appendix \ref{sec:spin_wave_approximation}, this exponent is given by $\eta_\mathrm{2D}(T)=T/(2\pi)$.
At the critical temperature, it reaches a value of
\begin{equation}
\eta_\mathrm{2D}(T_c) = \frac{1}{4}.
\end{equation}
Equations \eqref{M_2D_def} and \eqref{spin_correlation_2D} imply that the magnetization follows a power law decay:
\begin{equation}
M \sim L^{-\eta_\mathrm{2D}(T)/2}.
\end{equation}
In Fig.~\ref{fig_helicity_modulus}(c), we show $M$ as a function of the temperature for several system sizes.
The inset of Fig.~\ref{fig_helicity_modulus}(c) shows the relation of $M$ with $L$ at the critical temperature $T=0.9$.
This relationship fits well with a power law function described as $aL^{-b}$. 
The resulting exponent, $b=0.122$, closely matches the theoretical value $\eta_\mathrm{2D}(T_c)/2=1/8$.

\subsection{3D DRFXYM}

Now, we turn our attention to the 3D DRFXYM at zero temperature.
Similar to the 2D XY model, we evaluate the helicity modulus by measuring the response force to the twist of the spins.
Suppose twisting the spins at the top boundary of a 3D cubic simulation box of length $L$ by a slight angle $\theta$ with respect to the bottom boundary.
The helicity modulus $\Gamma$ is then expressed as:
\begin{equation}
\Gamma = \lim_{\theta \to 0} \frac{1}{L\theta} \left\langle \frac{\partial H}{\partial \theta} \right\rangle,
\label{Gamma_def}
\end{equation}
where $H$ is the Hamiltonian of the 3D DRFXYM.
Importantly, Eq.~\eqref{Gamma_def} can be applied to nonequilibrium situations, considering $\langle ... \rangle$ as the average for nonequilibrium steady states.
Assuming that the concept of dimensional reduction is valid, the universal jump relation is obtained by substituting the effective temperature Eq.~\eqref{effective_temperature} into Eq.~\eqref{hm_jump_2D}.
The resulting equation is
\begin{equation}
\frac{\Gamma_c v}{\Delta_c^2} = \frac{1}{\pi},
\label{hm_jump_3D}
\end{equation}
where $\Delta_c$ is the critical disorder strength.

In simulations, the equation of motion is integrated using the Euler method for $10^5$ steps, with a time discretization of $dt = 1/ 10v$, where $v$ represents the driving velocity.
We compute the time average for the helicity modulus $\Gamma$ after an initial transient phase, and it is also supplemented by ensemble averages over 50 to 100 different initial states.
For more details on the simulation of the DRFXYM, refer to Appendix \ref{sec:simulation}.
In Fig.~\ref{fig_helicity_modulus}(d), we show $\Gamma$ as a function of the disorder $\Delta$ for various system sizes $L$.
The behavior observed here mirrors that of the standard 2D XY model.
Specifically, the value of $\Gamma$ remains unaffected by changes in $L$ when $\Delta<\Delta_c \simeq 2.2$. 
However, for $\Delta>\Delta_c$, $\Gamma$ diminishes to zero as $L$ increases.
The solid curve represents $\Gamma = \Delta^2/(v \pi)$.
The point where this curve intersects with the $\Gamma$ curve close to $\Delta_c$ suggests that our universal jump relation in Eq.~\eqref{hm_jump_3D} is a good approximation.

The finite-size scaling can be performed similarly to how we did with the 2D XY model.
From Eqs.~\eqref{effective_temperature} and \eqref{xi_2D}, the correlation length $\xi(\Delta)$ is expected to behave as
\begin{equation}
\xi(\Delta) \propto \exp[B/(\Delta^2-\Delta_c^2)^{1/2}],
\label{xi_3D}
\end{equation}
where $B$ is a non-universal constant.
In Fig.~\ref{fig_helicity_modulus}(e), we plot $\Gamma$ against $\log[L/\xi(\Delta)]=\log L - B/(\Delta^2-\Delta_c^2)^{1/2}$.
We can observe data collapse with parameters $\Delta_c = 2.17$ and $B = 8.06$.

For weak disorder, the spin correlation function shows a power-law decay with anisotropic exponents \cite{Haga-15, Haga-18},
\begin{equation}
\langle \boldsymbol{\phi}(\boldsymbol{r}) \cdot \boldsymbol{\phi}(\boldsymbol{r}') \rangle \propto
\begin{cases}
		|\boldsymbol{r}-\boldsymbol{r}'|^{-\eta_\perp(\Delta)} & (\boldsymbol{r}-\boldsymbol{r}' \perp \boldsymbol{v}), \\
		|\boldsymbol{r}-\boldsymbol{r}'|^{-\eta_\parallel(\Delta)} & (\boldsymbol{r}-\boldsymbol{r}' \parallel \boldsymbol{v}),
\end{cases}
\label{spin_correlation_3D}
\end{equation}
where $\boldsymbol{v}=(v, 0, 0)$ is parallel to $x$-direction.
The derivation of Eq.~\eqref{spin_correlation_3D} with the spin-wave approximation is presented in Appendix \ref{sec:spin_wave_approximation}.
According to the dimensional reduction conjecture, the transverse direction of the driven disordered system corresponds to the spatial direction of the lower-dimensional counterpart (see Fig.~\ref{fig_DR_concept}).
Thus, from Eqs.~\eqref{effective_temperature}, the transverse exponent $\eta_\perp(\Delta)$ is given by
\begin{equation}
\eta_\perp(\Delta) = \eta_\mathrm{2D}\left( \frac{\Delta^2}{2v} \right)
\label{eta_perp_eta_2D}
\end{equation}
in terms of the exponent $\eta_\mathrm{2D}(T)$ for the 2D XY model.
Since the longitudinal direction is mapped to the time direction of the lower-dimensional system by the dimensional reduction, a simple scaling argument implies
\begin{equation}
\eta_\parallel(\Delta) = \frac{\eta_\perp(\Delta)}{2},
\label{eta_parallel_eta_perp}
\end{equation}
which is because Eq.~\eqref{steady_state_equation_2} contains the first-order derivative of time and the second-order derivative of space (see Appendix \ref{sec:spin_wave_approximation} for explicit calculations).

Let us consider the system size dependence of the magnetization,
\begin{equation}
M = \frac{1}{L^3} \left[ \int d^3 \mathbf{r} d^3 \mathbf{r}' \langle \boldsymbol{\phi}(\mathbf{r}) \cdot \boldsymbol{\phi}(\mathbf{r}') \rangle \right]^{1/2}.
\label{M_3D_def}
\end{equation}
Equations \eqref{spin_correlation_3D} and \eqref{M_3D_def} imply that, in the quasi-long-range ordered phase ($\Delta < \Delta_c$), the magnetization exhibits a power law decay,
\begin{equation}
M \sim L^{-\eta_\perp(\Delta)/2}.
\label{M_L_dependence_3D}
\end{equation}
Note that $\eta_\parallel(\Delta)$ does not influence Eq.~\eqref{M_L_dependence_3D} because, for sufficiently large $L$, the exponent of magnetization is determined by the largest exponents of the correlation function.
Thus, the dimensional reduction predicts that $M \sim L^{-1/8}$ at the transition point $\Delta = \Delta_c$.
In Fig.~\ref{fig_helicity_modulus}(f), we show the magnetization $M$ as a function of the disorder $\Delta$.
The inset shows $M$ as a function of $L$ at the tentative transition point $\Delta=2.2$.
The magnetization can be fitted by a power law function $aL^{-b}$ with an exponent $b=0.104$, which deviates approximately $20\%$ from the theoretical prediction of $1/8$.
This deviation might stem from the finite-size effect.
Since the correlation function given by Eq.~\eqref{spin_correlation_3D} features anisotropic exponents $\eta_\parallel(\Delta)<\eta_\perp(\Delta)$, for not sufficiently large $L$, the magnetization exponent may be slightly shifted towards the smaller $\eta_\parallel(\Delta)$, resulting in a smaller observed exponent relative to the theoretical expectation, which is anticipated in the limit as $L \to \infty$.
If we determine the transition point from $M \sim L^{-1/8}$, we get a slightly larger critical value, $\Delta_c=2.32$.

To verify the universal jump relation of the helicity modulus given in Eq.~\eqref{hm_jump_3D}, we determine the critical disorder $\Delta_c$ as follows.
Firstly, we compute the magnetization $M(\Delta, L)$ across various values of $\Delta$ and $L$, fitting the results to a power-law form $aL^{-b}$.
The critical disorder $\Delta_c$ is identified using the criterion $b=1/8$.
We show $\Delta_c$ as a function of the driving velocity $v$ in Fig.~\ref{fig_hm_jump}(a).
The solid line in Fig.~\ref{fig_hm_jump}(a) represents the theoretical prediction given by Eq.~\eqref{Delta_c_DR}.
As shown in Fig.~\ref{fig_hm_jump}(b), the deviation from the theoretical value decreases as $v^{-1/2}$.
Figure \ref{fig_hm_jump}(c) shows $\Delta_c^2/\Gamma_c v$ as a function of the driving velocity $v$, where $\Gamma_c$ is the jump of the helicity modulus at the transition point.
The determination method for $\Gamma_c$ is described in Appendix \ref{sec:simulation}.
Figure \ref{fig_hm_jump}(d) demonstrates that $\Delta_c^2/\Gamma_c v$ approaches $\pi$ as $v$ increases.
These findings are consistent with Eq.~\eqref{hm_jump_3D}.

\begin{figure*}
	\centering
	\includegraphics[width=\textwidth]{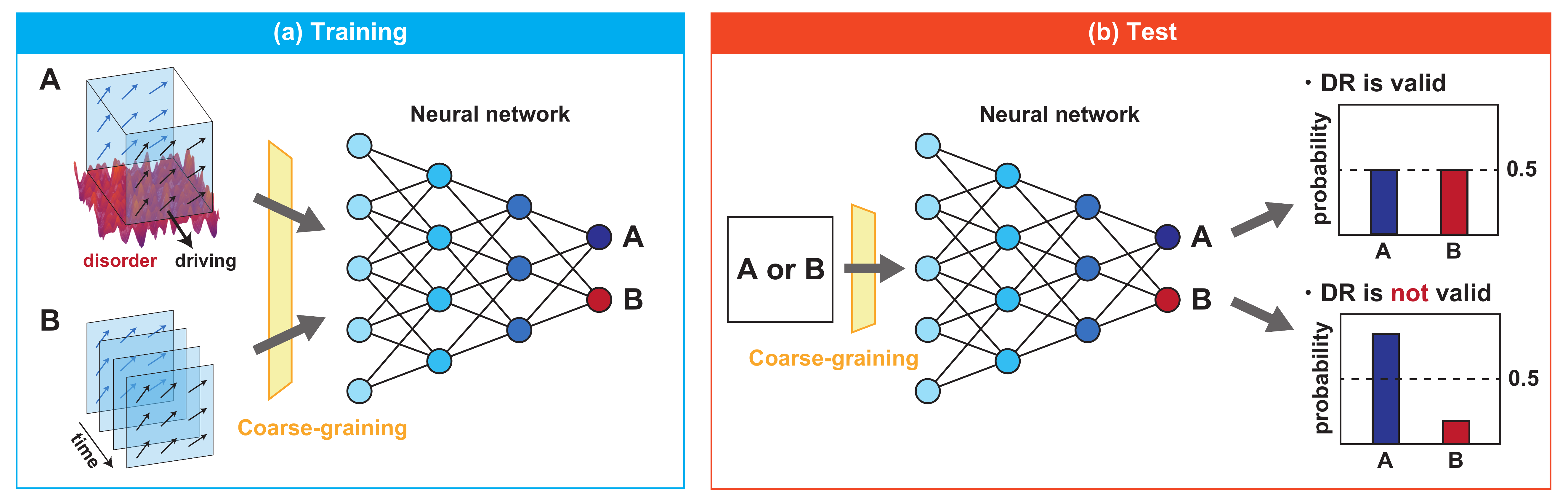}
	\caption{Illustration of the testing procedure for the dimensional reduction using a neural network.
		(a) Training Phase: During this stage, the neural network with two output neurons is trained using snapshots of the 3D DRFXYM and space-time trajectories of the 2D XY model.
		The objective is to differentiate between data from the 3D DRFXYM and the 2D XY model.
		(b) Testing Phase: In this stage, the trained network is employed to categorize new data from either the 3D DRFXYM or the 2D XY model. 
		The inability of the network to distinguish between the two indicates the validity of the dimensional reduction.
		A coarse-graining process is applied to the spin configurations before they are inputted into the neural network during both training and testing phases.}
	\label{fig_neural_network}
\end{figure*}

The deviations observed in Fig.~\ref{fig_hm_jump} from theoretical predictions may be attributed to the spatial discretization in the simulation procedure, detailed in Appendix \ref{sec:simulation}.
In these simulations, conducted within a moving frame, the spins rest while the random field moves at velocity $v$.
Given the spatial discretization $\Delta x$, each spin experiences a randomly varying field updated at time intervals $\Delta x/v$.
If the coupling between spins is neglected, the dynamics of each spin resemble those in the Euler-Maruyama method with time discretization $\Delta t = \Delta x/v$, effectively simulating a single spin influenced by white noise.
It is known that the (weak) error associated with the Euler-Maruyama method is $O(\Delta t)$ \cite{Higham-01}, proportional to $v^{-1}$ in the present situation.
This introduces a relative error in $\Delta_c$ proportional to $v^{-1}$, specifically, $(\Delta_c - \Delta_{c, \mathrm{DR}})/\Delta_{c, \mathrm{DR}} = O(v^{-1})$, or equivalently, $\Delta_c - \Delta_{c, \mathrm{DR}} = O(v^{-1/2})$, where the dimensional reduction value $\Delta_{c, \mathrm{DR}}$ is given by Eq.~\eqref{Delta_c_DR}.
Consequently, the error in $\Delta_c$ being proportional to $v^{-1/2}$ results in a corresponding error in the critical helicity modulus $\Gamma_c - \Gamma_{c, \mathrm{DR}}=O(v^{-1/2})$, assuming that $d\Gamma/d\Delta = O(1)$ near $\Delta=\Delta_c$.
This argument suggests that $\Delta_c^2/\Gamma_c v - \pi = O(v^{-1/2})$, which is consistent with Fig.~\ref{fig_hm_jump}(d).

\section{Test of dimensional reduction}
\label{sec:test_of_dimensional_reduction}

We employ a machine learning technique, specifically neural networks, to directly test the dimensional reduction conjecture using spin configurations. 
Neural networks excel at identifying intricate patterns in data, often beyond human perceptual abilities. 
Our task involves distinguishing a snapshot of the 3D DRFXYM from a space-time trajectory of the 2D XY model, which, according to the conjecture, should be indistinguishable.

The procedure is illustrated in Fig.~\ref{fig_neural_network}.
We use a multi-layer neural network with two output neurons, labeled A and B, tasked with a binary classification.
During the training phase [see Fig.~\ref{fig_neural_network}(a)], the network learns to output a ``1" from neuron A and a ``0" from neuron B when given a snapshot of the 3D DRFXYM. 
Conversely, for a space-time trajectory of the 2D XY model, neuron A outputs ``0" and neuron B outputs ``1".
In the testing phase [see Fig.~\ref{fig_neural_network}(b)], we apply the trained network to new, unseen data from either snapshots of the 3D DRFXYM or space-time trajectories of the 2D XY model. 
If the dimensional reduction conjecture is valid, the network should struggle to differentiate between the two, leading to a mere $50\%$ accuracy, equivalent to random chance or a coin toss. 
However, if the conjecture does not hold, the network should confidently and accurately distinguish between them.

\begin{figure}
	\centering
	\includegraphics[width=8.6cm]{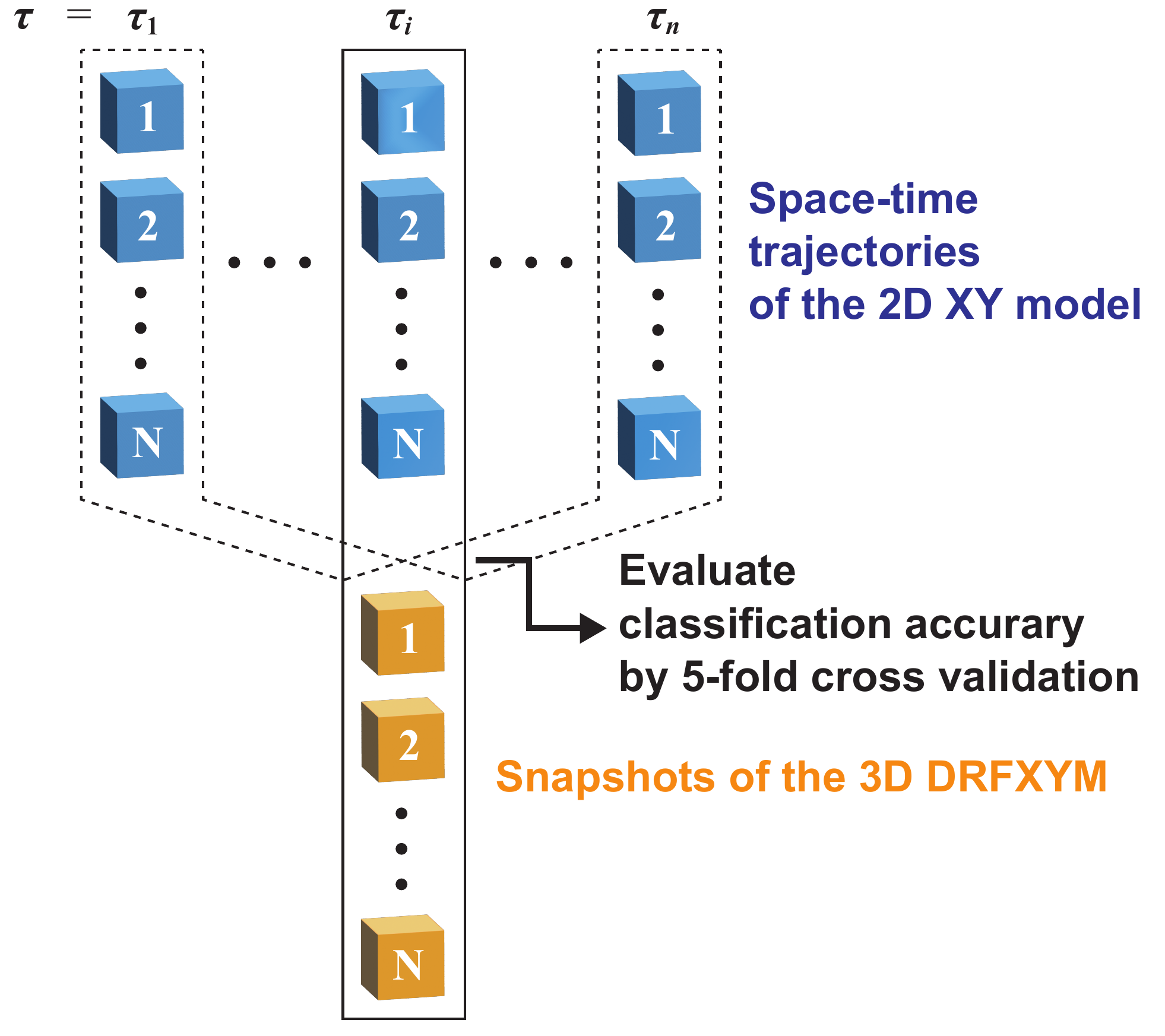}
	\caption{Composition of the training dataset.
		For each value of $\tau$, $N$ space-time trajectories of the 2D XY model are generated by solving Eq.~\eqref{2D_XY_dynamics_lattice}.
		These trajectories, along with $N$ snapshots of the 3D DRFXYM, comprise the training dataset.
		The classification accuracy is evaluated by 5-fold cross validation.}
	\label{fig_dataset}
\end{figure}

\begin{figure*}
	\centering
	\includegraphics[width=\textwidth]{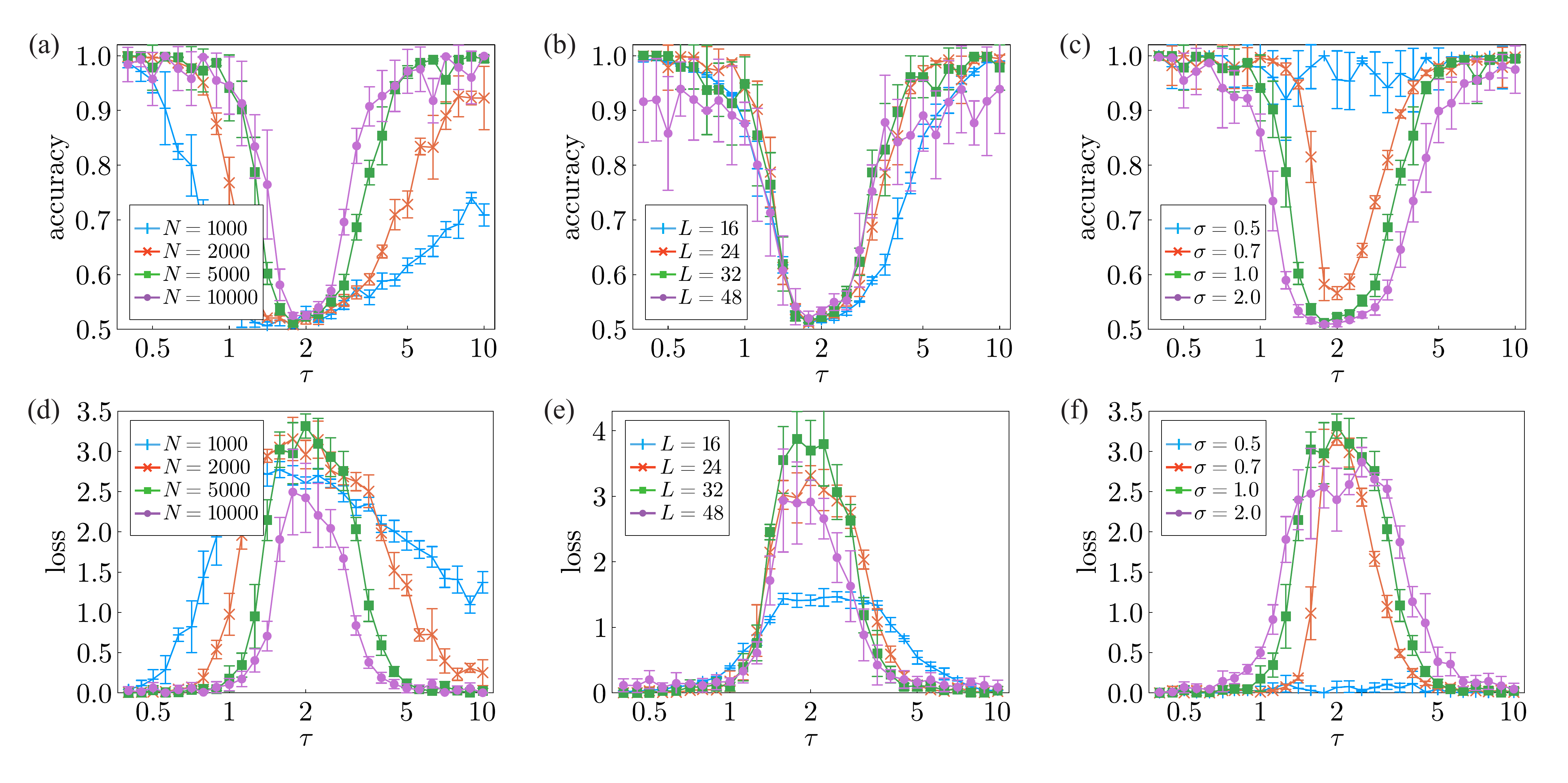}
	\caption{Accuracy [panels (a)-(c)] and loss [panels (d)-(f)] of the CNN in the testing phase. 
		The horizontal axis denotes the timescale parameter $\tau$ of the 2D XY model, shown in log-scale. 
		(a) and (d): Accuracy and loss for different numbers of samples.
		The system size is $L=24$ and the coarse-graining scale is $\sigma=1$. 
		(b) and (e): Accuracy and loss for different system sizes.
		The sample number is $N=5000$ and the coarse-graining scale is $\sigma=1$.
		(c) and (f): Accuracy and loss for different coarse-graining scales.
		The system size is $L=24$ and the sample number is $N=5000$.
		For large coarse-graining scales $\sigma \geq 1$, the accuracy reaches $0.5$ around $\tau=2$, which implies that the neural network fails to differentiate snapshots of the 3D DRFXYM from space-time trajectories of the 2D XY model.}
	\label{fig_DR_CNN}
\end{figure*}

We here explain how to generate the space-time trajectories of the 2D XY model.
The dynamics of the 2D XY model is described by
\begin{equation}
\tau \partial_t \phi_i(\boldsymbol{r}, t) = \left[ \nabla^2 - U'(\rho(\boldsymbol{r}, t)) \right] \phi_i(\boldsymbol{r}, t) + \xi_i(\boldsymbol{r}, t),
\label{2D_XY_dynamics}
\end{equation}
where $\boldsymbol{r}$ represents the 2D coordinates.
Here, we have introduced a control parameter $\tau$, which is related to the relaxation timescale.
The thermal noise $\xi_i(\boldsymbol{r}, t)$ satisfies
\begin{equation}
\langle \xi_i(\boldsymbol{r}, t) \xi_j(\boldsymbol{r}', t') \rangle = 2 \tau T_\mathrm{2D} \delta_{ij} \delta(\boldsymbol{r}-\boldsymbol{r}') \delta(t - t').
\end{equation}
In this context, the temperature $T_\mathrm{2D}$ should not be confused with $T$ in Eq.~\eqref{thermal_noise_xi}, which is set to zero when simulating the 3D DRFXYM.
From Eq.~\eqref{steady_state_equation_2}, the dimensional reduction conjecture predicts that snapshots of the 3D DRFXYM with a driving velocity $v$ are equivalent to space-time trajectories of the 2D XY model when $\tau=v$.
It is important to note that the timescale parameter $\tau$ does not influence the thermal equilibrium of the 2D XY model.
Instead of the continuous model given by Eq.~\eqref{2D_XY_dynamics}, it is convenient to use the conventional XY model on a square lattice.
Its time evolution is described by
\begin{equation}
\tau \partial_t \theta_i(t) = \sum_{j \in \langle i \rangle} \sin \left[ \theta_j(t) - \theta_i(t) \right] + \zeta_i(t),
\label{2D_XY_dynamics_lattice}
\end{equation}
where $\theta_i(t)$ is the spin angle at the $i$th site and $\langle i \rangle$ denotes the set of neighboring sites of the $i$th site.
The thermal noise $\zeta_i(t)$ satisfies
\begin{equation}
\langle \zeta_i(t) \zeta_j(t') \rangle = 2 \tau T_\mathrm{2D} \delta_{ij} \delta(t - t').
\label{thermal_noise_xi_2D}
\end{equation}
We solve Eq.~\eqref{2D_XY_dynamics_lattice} by using the Euler-Maruyama method with time discretization $dt=0.005$ to generate space-time trajectories of spin configuration $\{(\cos \theta_i(t), \sin \theta_i(t)) \}$ of the 2D XY model in thermal equilibrium.

It should be noted that the dimensional reduction conjecture predicts the equivalence of the \textit{large-scale} behaviors of the 3D DRFXYM and the 2D XY model, as indicated by Eq.~\eqref{steady_state_equation_2}.
Therefore, prior to inputting spin configurations into the neural network, we implement a coarse-graining process.
For any given 3D spin configuration, either a snapshot of the 3D DRFXYM or a space-time trajectory of the 2D XY model, a convolution procedure utilizing a Gaussian kernel with standard derivation $\sigma$ is performed (see Appendix \ref{sec:coarse_graining_procedure} for details).
This convolution procedure smooths the spin configurations by averaging over a neighborhood defined by $\sigma$.
The effect is to blur and reduce local fluctuations, emphasizing larger-scale structures in the data.
Subsequently, normalization standardizes the spin amplitude uniformly.
As we will demonstrate, this coarse-graining step is crucial for successfully applying the dimensional reduction.
Furthermore, it is confirmed that the specifics of the coarse-graining procedure do not qualitatively affect the trend of classification accuracy, as discussed in Appendix \ref{sec:coarse_graining_procedure}.
Additionally, to prevent any bias in the spin angle, a uniform global random phase in the range $[0, 2\pi]$ is added to all spins in each sample.

The structure of the neural networks employed in our study is detailed as follows. We utilize two different network architectures: a deep convolutional neural network (CNN) and a fully-connected neural network (FNN) with one hidden layer.
\begin{enumerate}
\item CNN: The input for the CNN consists of $L \times L \times L$ spin configurations, each with two channels representing the first and second components of $\boldsymbol{\phi}$.
The first layer is a 3D convolutional layer, featuring filters of size $4 \times 4 \times 4$ with a stride of 2, where the channel number increases to 8. 
The activation function employed here is the rectified linear (ReLU) function.
The second layer is the same convolutional layer as the first but maintains the channel number.
The output from this layer is reshaped and fed into the third layer, a fully-connected layer with 16 ReLU output units. 
This output is then passed to a fourth layer consisting of 2 output units, where the softmax function is applied to determine classification probabilities. 
Note that the last layer does not involve an activation function prior to the application of the softmax function.

\item FNN: The input for the FNN is a $2L^3$-dimensional vector, obtained by flattening the spin configuration. 
The initial layer is a fully-connected configuration with 64 output units, activated by the ReLU function.
This output advances to a second layer with 2 output units, and classification probabilities are again derived using the softmax function. 
Similarly to the CNN, the last layer in this network also does not include an activation function prior to softmax.
\end{enumerate}
In both models, the loss function is given by the cross-entropy between the predicted probability and the one-hot encoded target labels: $(1, 0)$ for the 3D DRFXYM and $(0, 1)$ for the 2D XY model.

Figure \ref{fig_dataset} depicts the composition of the training dataset.
The parameters of both the 3D DRFXYM and the 2D XY model are set to values corresponding to the transition point.
Specifically, we have $v=2$ and $\Delta=2.2$ for the 3D DRFXYM and $T_{\mathrm{2D}}=0.9$ for the 2D XY model.
For each value of the control parameter $\tau$, $N$ spin configurations are prepared for both the 3D DRFXYM and the 2D XY model, making a total of $2N$ samples.
Classification accuracy is assessed through 5-fold cross-validation, where in each fold, $8N/5$ samples are utilized for training and $2N/5$ samples are employed to estimate classification accuracy.
This process is iteratively performed across different values of $\tau$.
The network is implemented using the \texttt{Flux} library in the \texttt{Julia} programming language. 
Network parameters are optimized using the \texttt{Adam} optimizer with a learning rate of $10^{-3}$, and training is conducted over 30 epochs with a batch size of 50.

In Fig.~\ref{fig_DR_CNN}, the classification accuracy and loss function values for the CNN are plotted as functions of the timescale parameter $\tau$ for the 2D XY model. 
The estimation of accuracy and loss involves repeating five sets of 5-fold cross-validation, each with different random number seeds.
The average and standard deviation of both accuracy and loss are calculated, with the latter being indicated by error bars in Fig.~\ref{fig_DR_CNN}.
Panels (a) and (d) display the accuracy and loss for varying sample numbers with a system size $L=24$ and a coarse-graining scale $\sigma=1$.
Notably, the classification accuracy reaches $0.5$ around $\tau=v=2$, suggesting that the neural network is unable to distinguish between snapshots of the 3D DRFXYM and space-time trajectories of the 2D XY model. 
Note that the value of the loss function peaks around $\tau=v$ [see panel (d)], which serves as additional evidence of the neural network's inability to discern the differences between the two types of data.
While the dip in accuracy sharpens with increasing the sample number $N$, its depth does not change.
This suggests that the two types of data remain indistinguishable at $\tau=v$, even as $N \to \infty$.

Figures \ref{fig_DR_CNN}(b) and (e) show the accuracy and loss for varying system sizes with a sample number $N=5000$ and a coarse-graining scale $\sigma=1$.
As the system size $L$ increases, the accuracy profile converges towards a characteristic curve around $\tau=v$, with a minimum at $\tau=v$ reaching an accuracy of $0.5$.
It is observed that for larger system sizes (e.g., $L=48$), the accuracy diminishes at both small and large values of $\tau$.
This reduction in accuracy is attributed to networks are trapped in local minima within the parameter space during the training phase, which are associated with higher loss values (see Fig.~\ref{fig_appendix_train_loss} in Appendix \ref{sec:reproducibility}).
This explains the pronounced fluctuations in accuracy observed at the extremities of the $\tau$ range.
Additionally, the likelihood of training failures escalates with increasing system size $L$.

Finally, Figs.~\ref{fig_DR_CNN}(c) and (f) show the accuracy and loss for varying coarse-graining scales with a system size $L=24$ and a sample number $N=5000$.
A notable observation is that when the neural network is trained on configurations with a low coarse-graining scale, $\sigma=0.5$, it successfully distinguishes between snapshots of the 3D DRFXYM and space-time trajectories of the 2D XY model, even near $\tau=v$.
As $\sigma$ increases, the accuracy decreases towards $0.5$ around $\tau=v$. 
This trend suggests that the dimensional reduction may not be applicable for small-scale fluctuations in the system's configurations.

Note that the maximum value of loss around $\tau=v$ is larger than $-\log(0.5)=0.693$, the value of cross-entropy per sample if the network were to output $(0.5, 0.5)$ for all input data.
The large loss value means that the output of the neural network is highly biased, e.g., $(0.99, 0.01)$, but is uncorrelated with the target label, as indicated by the low accuracy.
In other words, the neural network is making incorrect decisions with a high degree of confidence.

\begin{figure*}
	\centering
	\includegraphics[width=\textwidth]{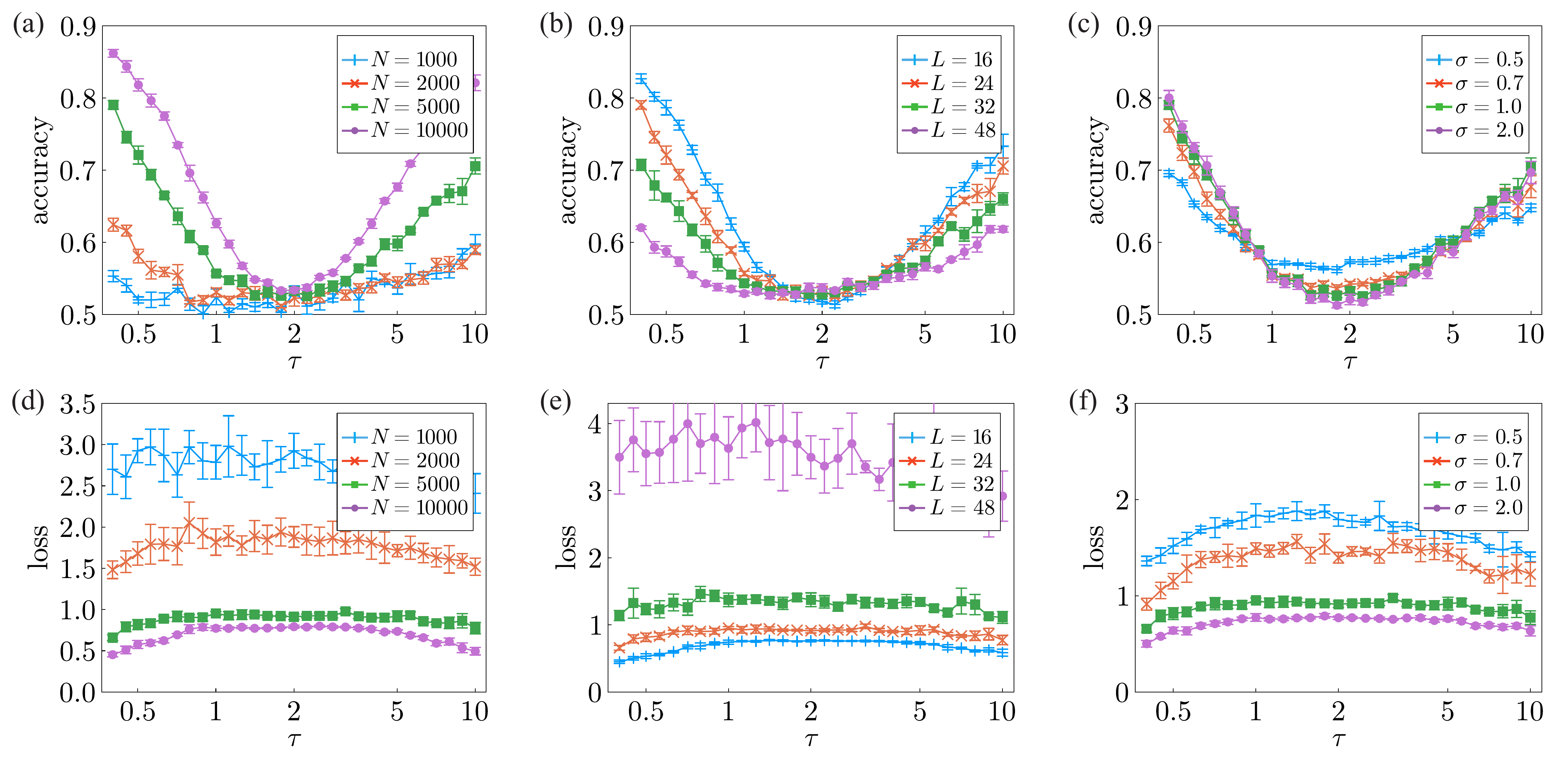}
	\caption{Accuracy [panels (a)-(c)] and loss [panels (d)-(f)] of the FNN in the testing phase. 
		The horizontal axis denotes the timescale parameter $\tau$ of the 2D XY model, shown in log-scale. 
		(a) and (d): Accuracy and loss for different numbers of samples.
		The system size is $L=24$ and the coarse-graining scale is $\sigma=1$. 
		(b) and (e): Accuracy and loss for different system sizes.
		The sample number is $N=5000$ and the coarse-graining scale is $\sigma=1$.
		(c) and (f): Accuracy and loss for different coarse-graining scales.
		The system size is $L=24$ and the sample number is $N=5000$.
		The qualitative behavior is the same as that observed in the CNN.}
	\label{fig_DR_FNN}
\end{figure*}

Next, let us consider the case of FNN.
We utilize a FNN with one hidden layer and have verified that adding more hidden layers does not improve the classification accuracy.
In Fig.~\ref{fig_DR_FNN}, the classification accuracy and loss function values for the FNN are plotted as functions of the timescale parameter $\tau$ for the 2D XY model.
The behavior aligns with that observed in the CNN, where accuracy approaches $0.5$ around $\tau=v=2$; however, the accuracy is lower in the FNN than in the CNN. 
This superiority of the CNN in phase classification has been highlighted previously in Ref.~\cite{Beach-18}.
Figures \ref{fig_DR_FNN}(a) and (d) display the accuracy and loss for varying sample numbers with a system size $L=24$ and a coarse-graining scale $\sigma=1$.
While the accuracy increases with the sample number $N$, its value at $\tau=v$ does not change, akin to the observations in Figs.~\ref{fig_DR_CNN}(a) and (d).
Figures \ref{fig_DR_FNN}(b) and (e) show the accuracy and loss for varying system sizes with a sample number $N=5000$ and a coarse-graining scale $\sigma=1$.
The reduction in accuracy with increasing system size $L$ is more pronounced than that observed for the CNN, a phenomenon also noted in Ref.~\cite{Beach-18}.
Finally, Figs.~\ref{fig_DR_FNN}(c) and (f) show the accuracy and loss for varying coarse-graining scales with a system size $L=24$ and a sample number $N=5000$.
In line with observations in the CNN, the accuracy at $\tau=v$ diminishes as the coarse-graining scale $\sigma$ increases.

\section{Conclusion}
\label{sec:conclusion}

We have investigated the BKT (Berezinskii--Kosterlitz--Thouless) transition in 3D driven disordered systems, employing machine learning techniques.
Central to this study is the recently proposed concept, the dimensional reduction conjecture.
It suggests that a static snapshot of a $D$-dimensional disordered system driven at a constant velocity is equal to a space-time trajectory of its $(D-1)$-dimensional pure counterpart.
Initially, we have examined the fundamental characteristics of the 3D BKT transition in the driven random field XY model (DRFXYM).
The universal jump relation \eqref{hm_jump_3D} of the helicity modulus within this model has been confirmed through nonequilibrium simulations.
Moving forward, we put the dimensional reduction conjecture to the test using machine learning. 
We trained neural networks to differentiate between snapshots of the 3D DRFXYM and space-time trajectories of the 2D XY model. 
When presented with new, unseen data from either the 3D DRFXYM or the 2D XY model, the neural network was unable to make accurate distinctions. 
This finding supports the dimensional reduction conjecture.

We discuss a potential experimental setup to realize the 3D DRFXYM.
A promising approach involves using nematic liquid crystals in porous media \cite{Araki-12}. 
In such a setup, the anchoring energy between the porous medium's surface and the liquid crystal's director promotes a random orientation of the director, effectively simulating a symmetry-breaking random field.
To align with the symmetry characteristics of the XY model, we can utilize the dielectric anisotropy of the liquid crystal. 
Specifically, when the liquid crystal exhibits negative dielectric anisotropy, e.g., $p$-azoxyanisole (PAA), its director tends to align perpendicular to an applied electric field \cite{deGennes}.
Consequently, liquid crystals in porous media under an external electric field can be roughly modeled by the random field XY model.
To drive the system out of equilibrium, we introduce a steady flow of liquid crystal through the porous medium by applying a pressure gradient.
Here, two aspects should be noted.
Firstly, in this setup, the influence of thermal fluctuations in the directors is minimal. 
Thus, the system can be approximated as the 3D DRFXYM at zero temperature.
Secondly, the directors of liquid crystals do not have distinct head and tail, resulting in a symmetry that is slightly different from the conventional XY model. 
Despite this, it is anticipated that such a difference will not significantly alter the qualitative nature of the phase transition.

Let us consider future possibilities and directions. 
As highlighted in the Appendix \ref{sec:failure_of_dimensional_reduction}, there are instances where the concept of dimensional reduction does not apply. 
The exact reasons and extent to which dimensional reduction fails are not fully understood \cite{Haga-17, Haga-19}.
In typical scenarios, the failure of dimensional reduction becomes apparent through macroscopic quantities, such as the order parameter and its two-point correlation function. 
These are quantities that can often be easily measured in experiments. 
However, there are potential situations where the breakdown of dimensional reduction is so subtle that it does not cause noticeable anomalies in these macroscopic quantities.
In such cases, the machine learning approach we have showcased in this study could prove to be a valuable tool. 
This technique may help detect minor disruptions in dimensional reduction that are only identifiable through snapshots of the system, offering insights that might not be evident through traditional methods.

\begin{acknowledgments}
	The author thanks Noriko Oikawa for valuable discussions.
	This work was supported by JSPS KAKENHI Grant Number JP22K13983.
\end{acknowledgments}

\appendix

\section{Spin-wave approximation}
\label{sec:spin_wave_approximation}

In this appendix, we overview the analysis of the spin-wave approximation, for comprehensive details, refer to Ref.~\cite{Haga-15}. 
It is crucial to note that this approximation is accurate primarily in the case of weak disorder, where the spin configuration $\boldsymbol{\phi}$ changes gradually in space, and the density of vortices is low. 
However, as vortices proliferate near the transition point, the spin-wave approximation loses its validity.

Let us start by examining the 2D XY model.
When vortices are absent, the magnitude of $\boldsymbol{\phi}$ becomes irrelevant.
Hence, $\boldsymbol{\phi}$ can be expressed as $\boldsymbol{\phi}=(\cos \theta, \sin \theta)$, with $\theta$ being the spin angle. 
The Hamiltonian for the standard XY model is then rewritten as
\begin{equation}
H = \frac{1}{2} \int d^2\boldsymbol{r} |\nabla \theta(\boldsymbol{r})|^2 = \frac{1}{2} \int \frac{d^2\boldsymbol{q}}{(2\pi)^2} q^2 \tilde{\theta}_{\boldsymbol{q}} \tilde{\theta}_{\boldsymbol{-q}},
\end{equation}
where $\tilde{\theta}_{\boldsymbol{q}}$ denotes the Fourier transform of $\theta(\boldsymbol{r})$.
The mean-square displacement of spins is calculated as
\begin{equation}
\langle [\theta(\boldsymbol{r}) - \theta(\boldsymbol{0})]^2 \rangle = T \int \frac{d^2\boldsymbol{q}}{(2\pi)^2} \frac{|1-e^{i \boldsymbol{q} \cdot \boldsymbol{r}}|^2}{q^2} = \frac{T}{\pi} \log r.
\end{equation}
Consequently, the spin correlation function is
\begin{equation}
\langle \boldsymbol{\phi}(\boldsymbol{r}) \cdot \boldsymbol{\phi}(\boldsymbol{0}) \rangle = e^{-\frac{1}{2} \langle [\theta(\boldsymbol{r}) - \theta(\boldsymbol{0})]^2 \rangle} \propto r^{-\eta_\mathrm{2D}(T)},
\end{equation}
with the exponent $\eta_\mathrm{2D}(T)$ given by
\begin{equation}
\eta_\mathrm{2D}(T) = \frac{T}{2\pi}.
\end{equation}
Note that this expression for $\eta_\mathrm{2D}(T)$ is only applicable at low temperatures.

Now, let us consider the 3D DRFXYM.
According to the steady state equation \eqref{steady_state_equation_1}, the spin angle $\theta(\boldsymbol{r})$ satisfies
\begin{equation}
v \partial_x \theta(\boldsymbol{r}) = \nabla^2 \theta(\boldsymbol{r}) - h_1(\boldsymbol{r}) \sin \theta(\boldsymbol{r}) + h_2(\boldsymbol{r}) \cos \theta(\boldsymbol{r}),
\end{equation}
where the last two terms on the right-hand side represent the force induced by the random field $(h_1(\boldsymbol{r}), h_2(\boldsymbol{r}))$.
The mean-square displacement of spins is calculated as
\begin{equation}
\langle [\theta(\boldsymbol{r}) - \theta(\boldsymbol{0})]^2 \rangle = 2 \Delta^2 \int \frac{d^3\boldsymbol{q}}{(2\pi)^3} \frac{1 - \cos \boldsymbol{q} \cdot \boldsymbol{r}}{q^4 + v^2 q_x^2},
\label{sw_appro_msd}
\end{equation}
where we have utilized approximations like
\begin{align}
&\langle h_i(\boldsymbol{r}) h_j(\boldsymbol{r}') \sin \theta(\boldsymbol{r}) \sin \theta(\boldsymbol{r}') \rangle \nonumber \\
&\simeq \langle h_i(\boldsymbol{r}) h_j(\boldsymbol{r}') \rangle \langle \sin \theta(\boldsymbol{r}) \sin \theta(\boldsymbol{r}') \rangle.
\end{align}
Equation \eqref{sw_appro_msd} is further evaluated to
\begin{equation}
\langle [\theta(\boldsymbol{r}) - \theta(\boldsymbol{0})]^2 \rangle  \propto
\begin{cases}
	\frac{\Delta^2}{2\pi v} \log r & (\boldsymbol{r} \perp \boldsymbol{v}), \\
	\frac{\Delta^2}{4\pi v} \log r & (\boldsymbol{r} \parallel \boldsymbol{v}),
\end{cases}
\end{equation}
leading to
\begin{equation}
\eta_\perp(\Delta) = \frac{\Delta^2}{4\pi v}, \quad \eta_\parallel(\Delta) = \frac{\Delta^2}{8\pi v}.
\end{equation}
These expressions align with Eqs.~\eqref{eta_perp_eta_2D} and \eqref{eta_parallel_eta_perp}.

\section{Simulation method for DRFXYM}
\label{sec:simulation}

In simulations, it is convenient to switch from the lab frame to a moving frame using the Galilei transformation: $\boldsymbol{r}' = \boldsymbol{r} - \boldsymbol{v}t$ and $\boldsymbol{\phi}'(\boldsymbol{r}')=\boldsymbol{\phi}(\boldsymbol{r})$.
Then, Eq.~\eqref{model_equation} transforms to
\begin{equation}
\partial_t \phi'_i(\boldsymbol{r}') = - \frac{\delta H[\boldsymbol{\phi}'(\boldsymbol{r}'); \boldsymbol{h}(\boldsymbol{r}' + \boldsymbol{v}t)] }{\delta \phi'_i(\boldsymbol{r}')},
\label{model_equation_moving_frame}
\end{equation}
with the temperature set to zero.
Our objective is to simulate the spin model with a random field that moves at a constant velocity.

Instead of the continuous Hamiltonian \eqref{Hamiltonian}, we use the conventional XY Hamiltonian on a cubic lattice:
\begin{equation}
H_1[\theta] = - \sum_j \sum_{\nu=x,y,z} \cos \left[\theta(\boldsymbol{R}_j) - \theta(\boldsymbol{R}_j + \boldsymbol{e}_\nu)\right],
\end{equation}
where $\theta(\boldsymbol{R}_j)$ is the angle of the spin at site $j$ with coordinates $\boldsymbol{R}_j=(m, n, l)$ $(m, n, l \in \mathbb{N})$, and $\boldsymbol{e}_\nu \:(\nu=x,y,z)$ is the unit vector along the $\nu$-direction.
The interaction between spins and the random field $\boldsymbol{h}(\boldsymbol{R}_j) = (h_1(\boldsymbol{R}_j), h_2(\boldsymbol{R}_j))$ is described by
\begin{equation}
H_2[\theta, \boldsymbol{h}] = - \sum_j \left[ h_1(\boldsymbol{R}_j) \cos \theta(\boldsymbol{R}_j) + h_2(\boldsymbol{R}_j) \sin \theta(\boldsymbol{R}_j) \right].
\end{equation}
Each component of $\boldsymbol{h}(\boldsymbol{R}_j)$ is sampled from a mean-zero Gaussian distribution with standard derivation $\Delta$.
The time evolution of $\theta$ is described by
\begin{equation}
\partial_t \theta(\boldsymbol{R}_j) = - \frac{\partial H_1[\theta] }{\partial \theta(\boldsymbol{R}_j)} - \frac{\partial H_2[\theta, \boldsymbol{h}_\mathrm{moving}(t)] }{\partial \theta(\boldsymbol{R}_j)}.
\label{EOM_simulation}
\end{equation}
The moving random field $\boldsymbol{h}_\mathrm{moving}$ is given by
\begin{equation}
\boldsymbol{h}_\mathrm{moving}(\boldsymbol{R}_j, t) = \boldsymbol{h}(\boldsymbol{R}_j + [vt] \boldsymbol{e}_x),
\end{equation}
where $[x]$ denotes the largest integer not greater than $x$.
A cubic simulation box with length $L$ is considered, with open boundary conditions along the $x$-direction.
The random field is generated at the front of the simulation box and shifts one step every $1/v$ time interval.
The equation of motion \eqref{EOM_simulation} is integrated using the Euler method with a time discretization of $dt = 1/ 10v$, meaning that 10 updates occur for each shift of the random field.
For the calculation of the magnetization, periodic boundary conditions are imposed along the $y$- and $z$-directions.

The components of the total magnetization are given by
\begin{equation}
\hat{\mathcal{M}}_x = \sum_j \cos \theta(\boldsymbol{R}_j), \quad \hat{\mathcal{M}}_y = \sum_j \sin \theta(\boldsymbol{R}_j).
\end{equation}
The averaged magnetization per spin is
\begin{equation}
M = \frac{1}{L^3} \left\langle \hat{\mathcal{M}}_x^2 + \hat{\mathcal{M}}_y^2 \right\rangle^{1/2},
\end{equation}
where $\langle ... \rangle$ denotes the long-time average in the steady state.

For the calculation of the helicity modulus, spins at the boundary at $y=L$ are twisted by an angle $\epsilon \ll 2\pi L$ relative to the boundary at $y=1$.
The response force against the twist is given by
\begin{equation}
f_\mathrm{twist}(\epsilon) = \frac{1}{L^2} \sum_j \sin \left[ \theta(\boldsymbol{R}_j + \boldsymbol{e}_y)-\theta(\boldsymbol{R}_j) \right],
\end{equation}
and the helicity modulus is
\begin{equation}
\Gamma = \frac{\langle f_\mathrm{twist}(\epsilon) \rangle}{\epsilon}.
\end{equation}
To avoid phase slip, $\epsilon$ should be less than $\pi$.
In main simulations, we set $\epsilon=\pi/2$.
Periodic boundary conditions are imposed along the $z$-direction.

To determine the critical disorder $\Delta_c$ and critical helicity modulus $\Gamma_c$ in Fig.~\ref{fig_hm_jump}, we first calculate the magnetizations $M(\Delta, L)$ for different system sizes $L=10, 15, 20, 30$, and they are fitted using a power-law form $L^{-\eta(\Delta)/2}$.
Once $\eta(\Delta)$ is obtained, the critical disorder $\Delta_c$ is determined by the condition $\eta(\Delta_c)=1/4$.
Next, we calculate the helicity modulus $\Gamma(\Delta, L)$ at $\Delta=\Delta_c$ for different system sizes $L=10, 15, 20, 30$.
As in the conventional XY model \cite{Weber-88}, the helicity modulus $\Gamma_c$ in the limit $L \to \infty$ is obtained by fitting $\Gamma(\Delta_c, L)$ with
\begin{equation}
\Gamma(\Delta_c, L) = \Gamma_c \left[ 1 + \frac{1}{2 \log L + C} \right],
\end{equation}
where $C$ is a non-universal constant.

\section{Coarse-graining procedure}
\label{sec:coarse_graining_procedure}

\begin{figure}
	\centering
	\includegraphics[width=8.6cm]{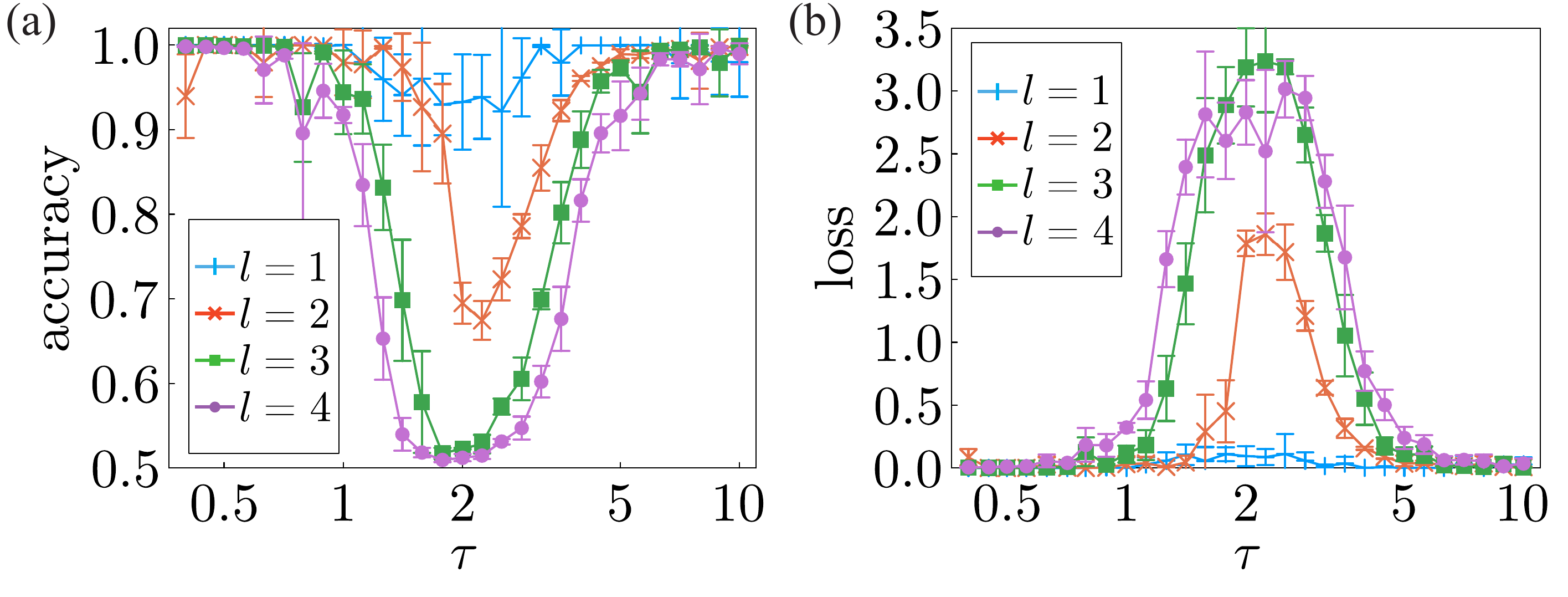}
	\caption{(a) Classification accuracy and (b) loss of the CNN employing a uniform kernel for the coarse-graining process. 
		The horizontal axis denotes the timescale parameter $\tau$ of the 2D XY model, shown in log-scale. 
		Kernel sizes are $l=1$ (no coarse-graining), $l=2$, $l=3$, and $l=4$.
		The system size is set at $L=24$ and the sample number at $N=5000$. 
		The observed trend is consistent with that of the coarse-graining process using a Gaussian kernel, as depicted in Fig.~\ref{fig_DR_CNN}(c).}
	\label{fig_appendix_coarse_graining}
\end{figure}

Given a three-dimensional scaler configuration $\phi(n_1, n_2, n_3)$ where $n_1, n_2, n_3 = 1,...,L$, we detail the coarse-graining procedure used in this study.
Let $K(m_1, m_2, m_3)$ with $m_1, m_2, m_3 = 1,...,l$ be a kernel with size $l$.
The convolution $\tilde{\phi}(n_1, n_2, n_3)$ is calculated as follows:
\begin{align}
\tilde{\phi}(n_1, n_2, n_3) =& \sum_{m_1, m_2, m_3=1}^l K(m_1, m_2, m_3) \phi \biggl(n_1+ \left\lfloor \frac{l+1}{2} \right\rfloor - m_1, \nonumber \\
&n_2+\left\lfloor \frac{l+1}{2} \right\rfloor-m_2, n_3+\left\lfloor \frac{l+1}{2} \right\rfloor-m_3 \biggr),
\end{align}
where $\lfloor x \rfloor$ denotes the largest integer not greater than $x$.
In the main text, we employ a Gaussian kernel:
\begin{align}
K(m_1, m_2, m_3) =& \ C_N \exp \biggl[ - \frac{1}{2\sigma^2} \left( m_1 - \frac{l+1}{2} \right)^2 \nonumber \\
&- \frac{1}{2\sigma^2} \left( m_2 - \frac{l+1}{2} \right)^2 - \frac{1}{2\sigma^2} \left( m_3 - \frac{l+1}{2} \right)^2 \biggr],
\end{align}
where $C_N$ is the normalization constant ensuring
\begin{equation}
\sum_{m_1, m_2, m_3=1}^l K(m_1, m_2, m_3) = 1.
\end{equation}
For a two-component spin configuration, this convolution is applied to each component, followed by a normalization procedure to standardize the amplitude of spins to unity.
Note that the coarse-graining transformation preserves the global $O(2)$ symmetry of the model.

To assess whether the choice of kernel affects the qualitative behavior of classification accuracy, we also evaluate a coarse-graining procedure using a uniform kernel $K(m_1, m_2, m_3)=1/l^3$, which is equivalent to a moving average over scale $l$.
Figure \ref{fig_appendix_coarse_graining} presents the classification accuracy and loss across different kernel sizes, showing a reduction in accuracy around $\tau=v$ with an increase in kernel size.
This behavior mirrors the results observed with a Gaussian kernel, as shown in Fig.~\ref{fig_DR_CNN}(c).

\section{Reproducibility of neural network results}
\label{sec:reproducibility}

\begin{figure}
	\centering
	\includegraphics[width=8.6cm]{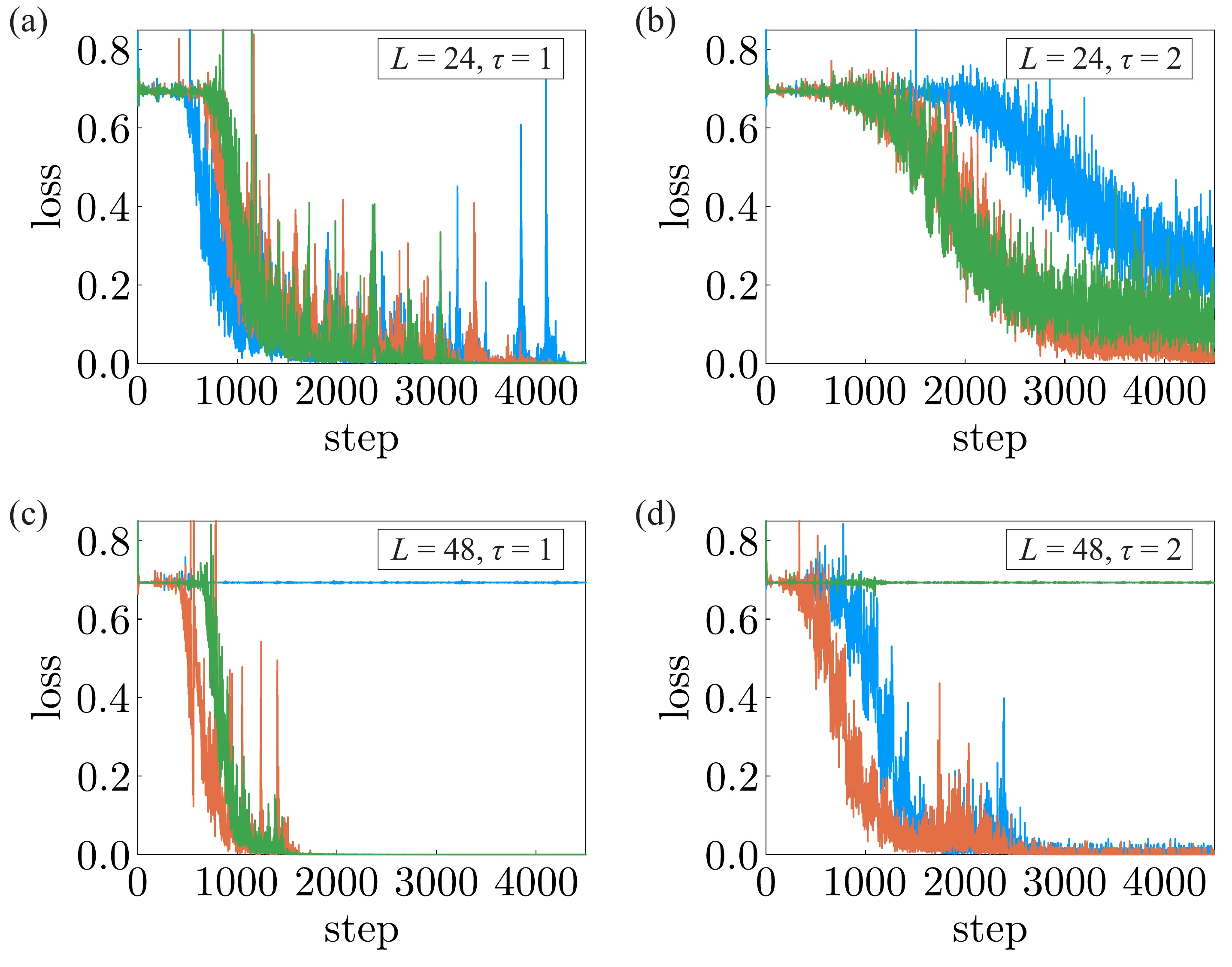}
	\caption{Evolution of the loss function of the CNN during the training phase for varying parameters: (a) $L=24, \tau=1$, (b) $L=24, \tau=2$, (c) $L=48, \tau=1$, (d) $L=48, \tau=2$, each with a sample number of $N=5000$.
		Three plots in each panel represent the loss values for three different random number seeds.
		At $\tau=2$, where the dimensional reduction is anticipated, the decline of the loss values becomes slow.
		Additionally, for $L=48$, some neural networks fail to reach minimal loss values due to entrapment in local minima of the loss function.}
	\label{fig_appendix_train_loss}
\end{figure}

In this Appendix, we examine the reproducibility of neural network outcomes with respect to variations in the random number seed, which influences data segmentation in cross-validation, the sequence of training data, and the initialization of network parameters.

Figure \ref{fig_appendix_train_loss} illustrates the evolution of the loss function during the training phase. 
The horizontal axis denotes each training step for the minibatches of data.
Three plots in each panel represent the loss values for three different random number seeds.
The initial loss value is approximately $-\log(0.5) = 0.693$, reflecting the cross-entropy per sample when the network predicts uniform probabilities of $(0.5, 0.5)$ for all inputs. 
At $\tau=1$, where dimensional reduction is not applicable, the loss rapidly declines to zero. 
Conversely, at $\tau=2$, where dimensional reduction is expected, the loss decreases more gradually, highlighting the challenge in distinguishing between the two types of data. 
Particularly for larger system size $L=48$, certain neural networks do not achieve minimal loss values due to being trapped in local minima of the loss function.

\begin{figure}
\centering
\includegraphics[width=8.6cm]{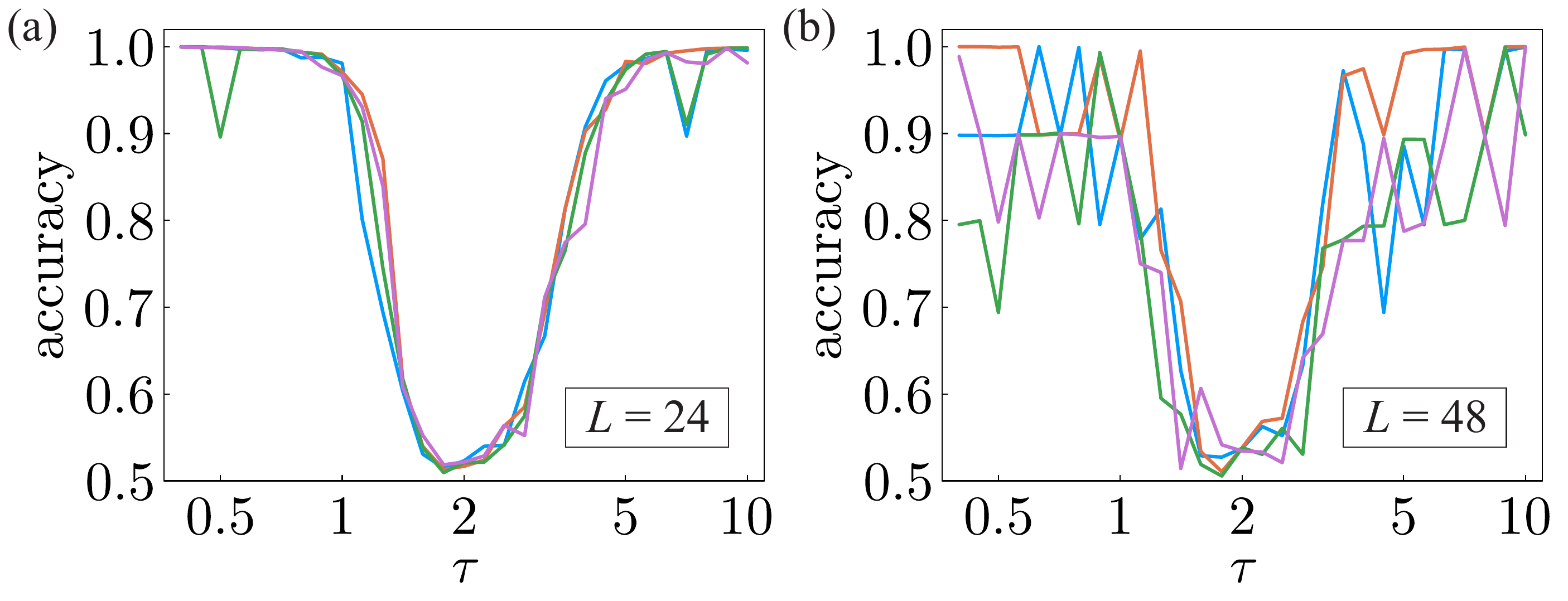}
\caption{Accuracy of the CNN from 5-fold cross-validation using four different random number seeds.
The horizontal axis denotes the timescale parameter $\tau$ of the 2D XY model, shown in log-scale. 
The system sizes are (a) $L=24$ and (b) $L=48$, each with a sample number of $N=5000$.
Notably, for $L=48$, significant fluctuations in accuracy are observed, attributable to some neural networks failing to achieve minimal loss values.}
\label{fig_appendix_accuracy_random_seed}
\end{figure}

Figure \ref{fig_appendix_accuracy_random_seed} displays the classification accuracy of the CNN obtained through 5-fold cross-validation using four different random number seeds.
It is important to note that Fig.~\ref{fig_DR_CNN} in the main text averages these accuracies across five different random number seeds.
For system size $L=48$, notable fluctuations in accuracy are observed, which are attributed to some neural networks' failure to achieve minimal loss values, as evidenced in Figs.~\ref{fig_appendix_train_loss}(c) and (d).

\section{Failure of the dimensional reduction for $N=1$}
\label{sec:failure_of_dimensional_reduction}

\begin{figure}[b]
	\centering
	\includegraphics[width=8.6cm]{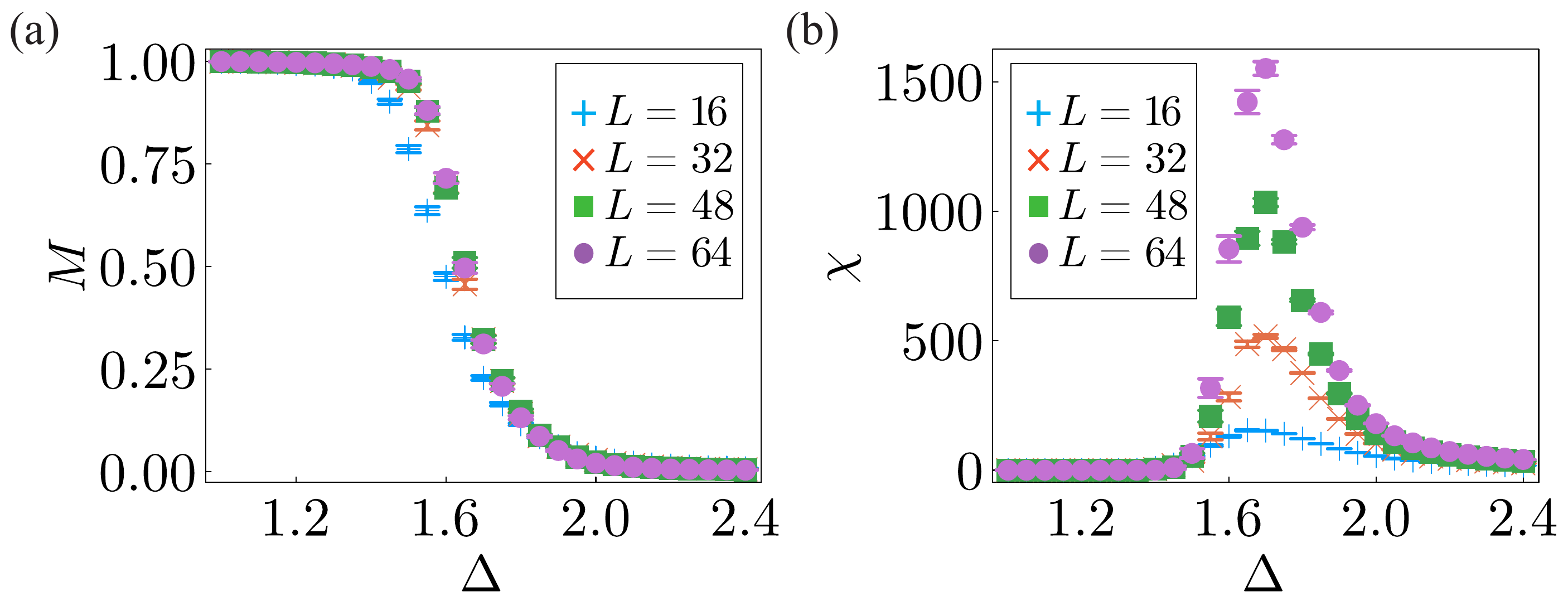}
	\caption{Magnetization $M$ [panel (a)] and its fluctuation $\chi$ [panel (b)] of the DRFIM as functions of the disorder strength $\Delta$.
		The system sizes are $L=16$, $32$, $48$, $64$, and the driving velocity is $v=4$.
		The peak of $\chi$ at $\Delta_c=1.7$ indicates the existence of a phase transition.}
	\label{fig_appendix_Ising}
\end{figure}

Let us consider the case of $N=1$ in the DRF$O(N)$M, i.e., the driven random field Ising model (DRFIM).
We aim to demonstrate that the 2D DRFIM, at zero temperature, undergoes a phase transition from a ferromagnetic phase to a paramagnetic phase as the disorder strength increases. 
This finding contradicts the dimensional reduction conjecture because it incorrectly predicts that the 2D DRFIM should behave like a one-dimensional pure Ising model, which does not exhibit any phase transition.

Instead of the continuous Hamiltonian \eqref{Hamiltonian}, we consider the conventional Ising model on a square lattice with length $L$:
\begin{equation}
H_1[S] = - \sum_j \sum_{\nu=x,y} S(\boldsymbol{R}_j) S(\boldsymbol{R}_j + \boldsymbol{e}_\nu),
\end{equation}
where $S(\boldsymbol{R}_j)=-1, 1$ represents the spin variable at site $j$ with coordinates $\boldsymbol{R}_j=(m, n)$ $(m, n \in \mathbb{N})$, and $\boldsymbol{e}_\nu \:(\nu=x,y)$ is the 2D unit vector along the $\nu$-direction.
The interaction between the spins and the random field $h(\boldsymbol{R}_j)$ is described by
\begin{equation}
H_2[S, h] = - \sum_j h(\boldsymbol{R}_j) S(\boldsymbol{R}_j),
\end{equation}
where $h(\boldsymbol{R}_j)$ is sampled from a mean-zero Gaussian distribution with standard derivation $\Delta$.

As detailed in Appendix \ref{sec:simulation}, the DRFIM is simulated in a moving frame with velocity $v$.
The Hamiltonian for the Ising model with a moving random field is given by
\begin{equation}
H[S, h, t] = H_1[S] + H_2[S, h_\mathrm{moving}(t)],
\end{equation}
\begin{equation}
h_\mathrm{moving}(\boldsymbol{R}_j, t) = h(\boldsymbol{R}_j + [vt] \boldsymbol{e}_x),
\end{equation}
where $[x]$ denotes the largest integer not greater than $x$.
The time evolution follows a standard Monte Carlo procedure at zero temperature.
In each step, a single spin is randomly selected for potential flipping.
If flipping the spin reduces $H[S, h, t]$, the update is accepted; if it increases $H[S, h, t]$, the update is rejected.
In one unit of time, $L^2$ updates are attempted.
Open boundary conditions are assumed along the $x$-direction, and periodic conditions along the $y$-direction.
The random field is generated at the front of the simulation box and shifts one step every $1/v$ time interval.

The total magnetization is given by
\begin{equation}
\mathcal{M} = \sum_j S(\boldsymbol{R}_j).
\end{equation}
For our analysis, we focus on two key observable quantities. 
The first is the average magnetization per spin,
\begin{equation}
M = \frac{1}{L^2} \langle \mathcal{M} \rangle,
\end{equation}
and the second is its fluctuation,
\begin{equation}
\chi = \frac{1}{L^2} \left[ \langle \mathcal{M}^2 \rangle - \left\langle \mathcal{M} \right\rangle^2 \right],
\end{equation}
where $\langle ... \rangle$ represents the long-time average in the steady state.
In thermal equilibrium, $\chi$ would equal the susceptibility of the magnetization in response to an external field. 
However, in our out-of-equilibrium model, this fluctuation-dissipation relationship generally does not apply.

Figure \ref{fig_appendix_Ising} shows the averaged magnetization $M$ and the fluctuation $\chi$ as functions of the disorder strength $\Delta$ for different system sizes.
In Fig.~\ref{fig_appendix_Ising}(a), we observe that the magnetization disappears when the disorder strength surpasses a critical value $\Delta_c \simeq 1.7$.
In Fig.~\ref{fig_appendix_Ising}(b), a peak in $\chi$ emerges at $\Delta=\Delta_c$ as the system size increases.
These observations support the existence of a phase transition from a ferromagnetic phase in $\Delta < \Delta_c$ to a paramagnetic phase in $\Delta > \Delta_c$.

The presence of long-range order in the 2D DRFIM can be understood through the following heuristic reasoning.
Firstly, it should be noted that the lower critical dimension of the random field Ising model in thermal equilibrium is two \cite{Imbrie-84, Bricmont-87}.
This is a significant finding because the equilibrium version of the dimensional reduction conjecture predicts the lower critical dimension to be three.
Secondly, introducing a constant velocity driving reduces the lower critical dimension by one.
This is because the advection term $v\partial_x \phi$ weakens the infrared singularity of the Green function from $\boldsymbol{k}^{-2}$ to $(\boldsymbol{k}^2 - iv k_x)^{-1}$ (see Appendix \ref{sec:spin_wave_approximation}).
Consequently, it is expected that the lower critical dimension for the DRFIM is one, implying the presence of long-range order in two dimensions.

\end{document}